\documentclass[apj]{emulateapj}

\newcommand{\myemail}{schenker@astro.caltech.edu}
\newcommand{\lt}{\ifmmode\,<\,\else \,$<$\,\fi}
\newcommand{\kms}{\ifmmode\,{\rm km}\,{\rm s}^{-1}\else km$\,$s$^{-1}$\fi}

\newcommand{\magarc}{\ifmmode {{{{\rm mag}~{\rm arcsec}}^{-2}}}
             \else {{{mag}$~${arcsec}$^{-2}$}}
             \fi}

\newcommand{\eg}{e.g.\@}

\newcommand{\Phik}{\Phi_{k}}
\newcommand{\MUV}{M_{\mathrm{UV}}}
\newcommand{\Nbin}{N_{\mathrm{bin}}}
\newcommand{\nexpk}{n_{\mathrm{exp},k}}
\newcommand{\nexpj}{n_{\mathrm{exp},j}}
\newcommand{\nobsk}{n_{\mathrm{obs},k}}
\newcommand{\nconk}{n_{\mathrm{con},k}}

\newcommand{\Veffk}{V_{\mathrm{eff}, k}}

\newcommand{\phiunit}{\mathrm{Mpc}^{-3}~\mathrm{mag}^{-1}}

\shorttitle{The UV Luminosity Function of Star-Forming Galaxies at z $\sim$ 7 and 8}

\begin{document}

\title{The UV Luminosity Function of Star-Forming Galaxies via Dropout Selection at Redshifts z $\sim$ 7 and 8 from the 2012 Ultra Deep Field Campaign}


\author {Matthew A Schenker\altaffilmark{1},
Brant E Robertson\altaffilmark{2},
Richard S Ellis\altaffilmark{1}
Yoshiaki Ono\altaffilmark{3},
Ross J McLure\altaffilmark{4},
James S Dunlop\altaffilmark{4},
Anton Koekemoer\altaffilmark{5},
Rebecca A A Bowler\altaffilmark{4},
Masami Ouchi\altaffilmark{3},
Emma Curtis-Lake\altaffilmark{4},
Alexander B Rogers\altaffilmark{4},
Evan Schneider\altaffilmark{2},
Stephane Charlot\altaffilmark{7},
Daniel P Stark\altaffilmark{3},
Steven R Furlanetto\altaffilmark{6},
Michele Cirasuolo\altaffilmark{4}
}

\altaffiltext{1}{Department of Astrophysics, California Institute of 
                Technology, MC 249-17, Pasadena, CA 91125; 
                \myemail}
\altaffiltext{2}{Department of Astronomy and Steward Observatory, University of Arizona, Tucson AZ 85721}
\altaffiltext{3}{Institute for Cosmic Ray Research, University of Tokyo,
                   Kashiwa City, Chiba 277-8582, Japan} 
\altaffiltext{4}{Institute for Astronomy, University of Edinburgh,
		Edinburgh EH9 3HJ, UK}
\altaffiltext{5}{Space Telescope Science Institute, Baltimore, MD 21218}
\altaffiltext{6}{Department of Physics \& Astronomy, University of California Los Angeles, Los Angeles CA 90095}
\altaffiltext{7}{UPMC-CNRS, UMR7095, Institut d'Astrophysique de Paris, F-75014, Paris, France}


\begin{abstract}

We present a catalog of high redshift star-forming galaxies selected to lie within the redshift range
$z\simeq$ 7-8 using the Ultra Deep Field 2012 (UDF12), the deepest near-infrared (near-IR) exposures 
yet taken with the Hubble Space Telescope. As a result of the increased near-infrared exposure time 
compared to previous HST imaging in this field, we probe $\sim$ 0.65 (0.25) mag fainter in absolute UV magnitude, 
at $z\sim$ 7 (8), which increases confidence in a measurement of the faint end slope of the galaxy luminosity function. 
Through a 0.7 mag deeper limit in the key F105W filter that encompasses or lies just longward
of the Lyman break, we also achieve a much-refined color-color selection that 
balances high redshift completeness and a low expected contamination fraction. We improve the 
number of drop-out selected UDF sources to 47 at $z\sim7$ and 27 at $z\sim8$. Incorporating brighter
archival and ground-based samples, we measure the $z\simeq$ 7 UV luminosity function to an absolute 
magnitude limit of $M_{\mathrm{UV}}=-17$ and find a faint end Schechter slope of $\alpha=-1.87^{+0.18}_{-0.17}$. Using a similar
color-color selection at $z\simeq$8 that takes account of our newly-added imaging in the F140W filter,
and incorporating archival data from the HIPPIES and BoRG campaigns, we provide a robust estimate
of the faint end slope at $z\simeq$8, $\alpha=-1.94^{+0.21}_{-0.24}$. We briefly discuss our results in the
context of earlier work and that derived using the same UDF12 data but with an independent photometric 
redshift technique (McLure et al 2012).

\end{abstract}

\keywords{cosmology: reionization --- galaxies: evolution --- galaxies: formation}

\section{Introduction}\label{sec:intro}

Great progress has been made in recent years in studies of the population of star-forming
galaxies at redshifts $z\simeq7-8$.  Following installation of the infrared Wide Field Camera 3 (WFC3), on the 
Hubble Space Telescope (HST), the number of candidates has risen from a few \citep{bouwens2008a} to 
$\simeq$ 100 \citep{mclure2010a,bouwens2010b,oesch2010a,yan2010a}.  In addition to providing hints
of the early galaxy population to $z\simeq8$, previous data sensitive to $z\sim7$ galaxies 
have provided initial determinations of their rest-frame UV colors,
stellar populations \citep{mclure2011a,bouwens2012b,dunlop2012a}, stellar 
masses and likely ages \citep{labbe2010a,gonzalez2010a,mclure2011a,finkelstein2012a}, and
nebular emission line strengths \citep{labbe2012a}. Our work builds upon these
previous efforts to present the first drop-out selected samples and luminosity function determinations for
redshift $z\sim7$ and $z\sim8$ sources from the 2012 Hubble Ultra Deep Field project (hereafter UDF12; 
GO 12498, PI: R. Ellis).

Before UDF12, progress has naturally been greatest at redshift $z \simeq 7$ where synergy between ground- and space-based
surveys has effectively exploited the full dynamic range of accessible galaxy luminosities. Early surveys 
from Subaru \citep{ouchi2009a}, and the ESO Very Large Telescope \citep{castellano2010a}, have probed the 
luminous component of the star-forming population over an area $>$ 1000 arcmin$^2$.  More recently,
the UltraVISTA survey has covered 3600 arcmin$^2$ in the COSMOS field, locating $z \simeq 7$ galaxies 
to $M_{\mathrm{UV}} = -22.7$ \citep{bowler2012a}.  

However, only HST can probe the important faint end of the galaxy luminosity function at these redshifts.
An early result from the 2009 Hubble Ultra Deep Field campaign (GO 11563, PI: Illingworth, hereafter UDF09)
was the discovery of an abundant population of sub-luminous galaxies at $z \simeq 7$ \citep{oesch2010a,bunker2010a,mclure2010a}
corresponding to a Schechter faint end slope  $\alpha$ between -1.7 to -2.0. In such a distribution, the bulk
of the integrated luminosity density arises from low luminosity galaxies that may be responsible to
maintaining cosmic reionization \citep{robertson2010a}. 

Clearly the luminosity function of star-forming galaxies at redshifts $z\simeq$ 7-8 is of great importance. However, given
the large range in luminosity that must be sampled, wider-field HST surveys have proved an important complement to 
panoramic ground-based surveys. WFC3 data from the GOODS Early Release Science (ERS) \citep{windhorst2011a}, 
and CANDELS fields \citep{grogin2011a,koekemoer2011a} have
sampled intermediate luminosities $-21\lesssim M_{\mathrm{UV}}\lesssim-19$.
The HIPPIES and BoRG pure parallel surveys have provided additional candidates at $z \sim 8$ \citep{yan2011a,trenti2011a,bradley2012a}.

Collectively these surveys have provided a reasonably clear view of the galaxy luminosity function at $z\simeq$7
at the luminous end but there remain disagreements regarding the precision with which the faint end slope is 
determined. While the UDF09 collaboration has measured a faint-end slope of $\alpha_{z\sim7} = -2.01\pm0.21$ 
\citep{bouwens2011a} incorporating the UDF, parallel fields, and the ERS data, a competing
determination utilizing the size-luminosity relation measured from the UDF09 data and 
the CANDELS Deep+Wide surveys in three fields finds a shallower faint-end slope of 
$\alpha_{z\sim7} = -1.7\pm0.1$ \citep{grazian2012a}.  The luminosity function at $z\simeq$8 is even more uncertain,
both because of the limited depth of the necessary photometry \citep{dunlop2012a} and 
the possibility of contamination from lower redshift sources.

Distant star-forming galaxies are commonly found using some variant of the Lyman break technique pioneered 
by \cite{steidel1996a}.  At $z \gtrsim 6.5$, the opacity due to neutral hydrogen in the intergalactic medium 
means the source flux below a rest-wavelength $\lambda_{rest} = 1216$\AA~is dimmed by factors $\geq 5$ \citep{madau1995a}.  
A search exploiting this effect has utilized one of two methods. In the `dropout' technique, objects are 
selected within a carefully-chosen window in color-color space specifically tuned to select star-forming galaxies 
within the required redshift range while minimizing the contribution from lower redshift contaminants.
At $z \sim 7$ (where the Lyman break falls near the overlap of the HST $z_{850}$ and $Y_{105}$ filters)
the Lyman dropout is chosen via a red color in $z - Y$, and the star-forming nature of the galaxy via a blue color in $Y - J$.
Additionally candidates are required to lie below a certain threshold in deep optical data; this further
limits contamination by lower redshift sources. Early demonstrations of the drop-out technique at redshifts $z\gtrsim 6$ were
presented by \cite{bunker2004a} and \cite{bouwens2004a}.

An alternative approach uses the full array of broadband detections and upper limits in the context of 
photometric redshift codes \citep{mclure2010a,finkelstein2010a,mclure2011a}.  These codes employ
a range of synthetic spectra for galaxies \citep[e.g.,][]{bruzual2003a} over all redshifts of interest and 
optimum fits are produced for each source in the catalog.  As in the dropout selection technique, leverage 
comes primarily from the Lyman break but the method is particularly advantageous when detections 
are available in more than 3 filters. 

In general, agreement between the two techniques is often quite good \citep{mclure2011a}.  However,
the methods are distinct and each has features relevant for interpreting their photometric samples.
The SED method provides redshift probability distributions for individual sources but may be susceptible
to systematic errors inherent in population synthesis models, such as uncertainties in the reddening law
and star-formation histories, when differentiating between low-redshift interlopers and true high-redshift
sources. In contrast, 
our drop-out selection only requires the observed color information independent of stellar population
synthesis modeling, but requires careful simulations to quantify the possible presence of low-redshift
contaminants satisfying the break criterion.  Given their complementarity and differing systematics,
independent luminosity function determinations from both methods will be helpful in furthering progress.
The goal of the present paper is therefore to exploit this unique data to provide the
best current constraints on the UV luminosity function of star forming galaxies at
redshift 7 and 8 using the dropout technique. A companion UDF12 
paper \citep{mclure2012a} presents the results of a search through the data
for $z \geq 6.5$ candidates using the photometric redshift technique.  

The present paper is one in a series devoted to scientific results from the UDF12 campaign,
which provides a significant advance over the earlier UDF09 imaging in this field appropriate
for the present luminosity function study and the complementary analysis discussed
in \cite{mclure2012a}. The UDF12 survey design and its data processing is discussed in 
\citet{koekemoer2012a}. Public versions of the final  reduced WFC3/IR UDF12 images, 
incorporating all earlier UDF data, are available to the community on the UDF12 web 
page\footnote{\rm http://udf12.arizona.edu/}.  Initial $z\ge8.5$ detections in the UDF12 data were presented
by \cite{ellis2012a}, while the UV continua of high-redshift sources were measured by 
\cite{dunlop2012b}. A review of the implications of the survey in the context of cosmic 
reionization including the results of the present paper is provided in \citet{robertson2012a}. 

A plan of the paper follows. In Section 2 we introduce the UDF12 data and the brighter
ground-based and HST data sets and their reductions essential for realizing an analysis
of the luminosity function at $z\simeq$7 and 8. Section 3 discusses the important
decisions we have taken in color-color space to optimize the completeness of
galaxies at these redshifts, while minimizing contamination from lower redshift sources.
Section 3 also presents the final list of galaxies in the two redshift intervals that we use
for our analysis. In Section 4 we present our luminosity functions and in Section 5
we briefly discuss our results in the context of earlier work, highlighting the important
advances made possible through the UDF12 campaign.

Throughout this paper, we adopt a $\Lambda-$dominated, flat universe with $\Omega_{\Lambda}=0.7$, 
$\Omega_{M}=0.3$ and $\rm{H_{0}}=70\,\rm{h_{70}}~{\rm km\,s}^{-1}\,{\rm Mpc}^{-1}$. 
All magnitudes in this paper are quoted in the AB system \citep{oke1983a}.  We will refer
to the HST ACS and WFC3/IR filters F435W, F600LP, F606W, F775W, F814W, F850LP, F105W, 
F098M, F125W, F140W, and F160W as $B_{435}$, $V_{600}$, $V_{606}$, $i_{775}$, $i_{814}$,
$z_{850}$, $Y_{098}$, $Y_{105}$, $J_{125}$, $J_{140}$, and $H_{160}$, respectively.

\section{Data}

Central to any analysis of the galaxy luminosity function is the collation of a complete
sample of galaxies within the chosen redshift interval spanning a wide range in 
luminosities, free from bias and with any interlopers minimized. In this section we introduce the UDF12
and more luminous auxiliary datasets.

\subsection{UDF}

To provide the best constraints on the faint end of the luminosity function at $z\simeq$ 7 and 8,
the primary advance of this paper is the increased depth and redshift fidelity
provided by our new UDF12 survey.
The UDF12 program dataset \citep[described in full in ][]{koekemoer2012a} represents a 
significant improvement over the previous UDF09 observations in several respects.
In particular, the survey was purposefully designed to improve our understanding of the
redshift $z\simeq7$ and 8 luminosity functions. Firstly, 
we increased the total exposure time in the key $Y_{105}$ filter over that in the UDF09
campaign by a factor of ~4, with the addition of 71 new orbits.  As this filter lies at the 
edge of the Lyman break for galaxies at $z \simeq 7$, this ensures we can probe significantly
fainter (by $\simeq$0.4 mag) in absolute magnitude at $z\simeq$7 and with considerably greater 
fidelity in redshift selection at $z\simeq$8 \citep{ellis2012a}. A further improvement is the addition of 
comparable imaging in the newly-utilized $J_{140}$ filter. By stacking our detections in this filter 
with either those at $J_{125}$ at $z\simeq$7 or $H_{160}$ at $z\simeq$8, we secure improved 
detections that correspond to extending the depth by an additional $\simeq$0.1 mag in each case.    

Our final analysis is based on the compilation discussed by \citet{koekemoer2012a} which 
incorporates all earlier WFC3 imaging in the UDF including the earlier UDF09 campaign
(GO 11563; PI: Illingworth) and less deep imaging undertaken as part of the CANDELS survey (GO 12060, PIs: Faber \&
Ferguson). In total the imaging constitutes 100, 39, 30, and 84 orbits in each of the
$Y_{105}$, $J_{125}$, $J_{140}$ and $H_{160}$ filters respectively (see \citealt{koekemoer2012a}
for further details). An important associated dataset in this field is the ultra deep ACS imaging
data from the 2004 UDF campaign \citep{beckwith2006a} essential for a further rejection
of low redshift sources.

\subsection{Auxiliary data}

To constrain the bright end of the $z\simeq$ 7 and 8 luminosity functions we take advantage of
several auxiliary WFC3 datasets which are somewhat shallower than our UDF12 data but nonetheless
unique in their coverage and depth. At $z\simeq7$, we include the UDF parallel fields, UDF-P1 and P2 
(sometimes referred to as UDF-P12 and P34, respectively) from various surveys including UDF05
(GO 10532, PI: Stiavelli) and the aforementioned UDF09 survey, the Early Release Science(ERS) WFC3 campaign
from the WFC3 Science Team (GO 11359, PI: O$'$Connell; \citealt{windhorst2011a}), and imaging in the GOODS-S field from  
CANDELS-Deep \citep{grogin2011a,koekemoer2011a}.  We also adopt the datapoints from the 
ground-based surveys of \cite{ouchi2009a}, \cite{castellano2010a} and \cite{bowler2012a}, which are
provide vital observations of the rare population of galaxies brighter than $M_{\mathrm{UV},\star}$.

At $z \sim 8$, it is difficult for ground-based programs to provide any meaningful constraints,
so we instead include data provided by two
HST pure-parallel WFC3 programs:  HIPPIES (GO/PAR 11702, PI: Yan; \citealt{yan2011a}) and BoRG (GO/PAR 11700, 
PI; Trenti; \citealt{trenti2011a,bradley2012a}).  As the shortest wavelength coverage of these surveys is provided only
with the $V_{606W}$ (or $V_{600LP}$), $Y_{098}$, and $J_{125}$ filters, a robust spectral break can only be verified
between $Y_{098}$ and $J_{125}$, as the wavelength spanned between $V_{606W}$ and $Y_{098}$ is too great and
there remains no optical rejection filter at shorter wavelengths.  Thus,
these fields can only usefully identify galaxies at $z \sim 8$, and therefore we do not use them for our selection at $z \sim 7$.

We summarize the filter coverage, survey area, limiting depths in the selection filter(s) and the number
of high redshift candidates for each of these auxiliary datasets in Table 1. 

\subsection{Data Reduction}

Prior to applying photometric color cuts optimized for the selection of $z\simeq$7 and 8 galaxies,
each survey data was similarly reduced to provide a series of processed and calibrated WFC3 frames.
We describe below the data reduction steps taken for each field.

\subsubsection{UDF and Parallels}

The preliminary processing stages that yield images ready for source selection are discussed in detail by 
\cite{koekemoer2012a}.  Briefly, we first process the raw images using the Pyraf/STDAS task \textit{calwf3},
which flags bad pixels, and corrects for bias and dark current throughout the detector.  After processing, the 
images are then registered and stacked using a version of the MultiDrizzle algorithm \citep{koekemoer2003a} 
to create the final mosaics on a pixel scale of 0.03$''$ per pixel.  This processing was carried out to create 
reductions of the UDF, UDF-P1, and UDF-P2 fields.  

\subsubsection{ERS and CANDELS-Deep}

For the ACS data in the CANDELS-Deep and ERS fields, we use the publicly available v2.0 mosaics from the 
GOODS campaign \citep{giavalisco2004a}.  We augment this with our own compilations of both the $i_{814}$
and $z_{850}$ data taken in parallel during the CANDELS WFC3 campaign.  We combine the publicly available
single epoch mosaics for these filters weighting by exposure time, using the image combination routine 
SWARP \citep{bertin2002a}.  In the case of the $z_{850}$ data, we add this to the already existing GOODS mosaic.  

We combine the public WFC3 mosaics from the CANDELS team \citep{koekemoer2011a} in the same
manner across CANDELS-Deep.  For the ERS WFC3 data, we use the reduction described in \citet{mclure2011a}.
Due to the lack of sub-pixel dithering available in the wider area fields, the CANDELS and ERS mosaics were 
produced with final pixel scales of 0.06$''$ per pixel.

\subsubsection{BoRG + HIPPIES}

Data from the BoRG and HIPPIES programs differ significantly from the prior data sets, as these are pure parallel
HST surveys.  Details of the observation strategy
can be found in \cite{trenti2011a} and \cite{bradley2012a}.  The complete dataset presented in \cite{bradley2012a} 
consists of 59 independent fields.  Due to the different nature of dealing with pure parallel data of limited depth, and
the survey's high expected contamination rate of $\sim 40 \%$, we make no effort to undertake an analysis ourselves.
Instead, we adopt the data points of \cite{bradley2012a}, and in Section 4, determine fits with and without these 
data added.

\subsection{Photometry}

In the case of the UDF12, which pushes HST to new limiting depths, we adopted a robust technique to locate 
and measure the fluxes from each faint source, rather than relying on errors output from the SExtractor source
extraction code \citep{casertano2000a}. As the correlated noise produced as a result of applying the 
Multidrizzle algorithm produces a subtle underestimate of uncertainties when using SExtractor,
in the specific case of the UDF12 we chose to use our own IDL photometry routine to 
compute all fluxes quoted in this paper. 

We briefly summarize the various stages. Processing proceeds by using a $\chi^2$ stack of all the images to 
identify regions of blank sky over the area in question.  A grid of blank apertures is then generated, with separation 
larger than the {\tt pixfrac} footprint of Multidrizzle, ensuring that noise between adjacent apertures is not correlated.  
To estimate the level of any residual background  around an object of interest, we take the median flux of the 50 
closest blank apertures.  We adopt the root mean square of the flux in these blank apertures as our uncertainty 
in the background level.  Fluxes are then computed using the APER routine in IDL \citep{landsman1993a}.
For the shallower non-UDF12
fields, although we utilized SExtractor to compute fluxes and background levels, the flux uncertainties were 
still estimated using this improved technique.

As the HST point spread function (PSF) varies significantly with wavelength (particularly between the ACS optical
images and near-infrared images from WFC3), it is important to account for this change when measuring accurate 
colors. To improve detections and color measurements for the faintest sources, aperture sizes should be quite small yet
properly account for wavelength-dependent PSF variations. We chose circular apertures whose diameters 
encircle $\geq 70\%$ of the total flux from a point source.  For the ACS $B_{435}$, $V_{606}$,  $i_{775}$, $ i_{814}$, 
and $z_{850}$ filters, we adopt aperture diameters of 0.30$''$  in all fields processed with 0.03$''$ pixel diameter 
(UDF12, UDF-P1, and UDF-P2), and 0.40$''$ in all other fields.  For the WFC3 filters, the aperture diameters are 
0.40$''$, 0.44$''$, 0.47$''$, and 0.50$''$, for $Y_{105}/Y_{098}$,  $J_{125}$, $J_{140}$, and $H_{160}$, respectively. 
Such small apertures remain meaningful both because of the FWHM of the HST PSF (ranging from 0.09$''$ in $B_{435}$ 
to 0.17$''$ in $H_{160}$), and the precise degree of alignment of the individual image sub-exposures (better the 0.005$''$ in the 
UDF; \citealt{koekemoer2012a}). In a related paper, \cite{ono2012a} validate these choices by  measuring half-light radii, 
r$_{hl}$, for stacks of our high-redshift samples, and find values of 0.36 kpc at both $z\sim7$ and $8$, which translate 
to angular half-light radii of $0.07''$ (not including any broadening due to the PSF).

\section{Candidate selection}

We now turn to the photometric selection of star-forming galaxies at $z\simeq$ 7 and 8, using
the photometric catalogs generated as discussed in Section 2. A key issue for our dropout selection
technique is the optimum choice of the two color cuts used to select candidates. The goal is to balance
completeness in high $z$ selection against a low fraction of foreground interlopers and spurious
sources. As the available filter sets differ for each of the component surveys, we discuss each case in turn.

\subsection{Potential contaminants}

A selection criterion tuned to select only high-redshift galaxies at high confidence,
with essentially no contamination, would be impractical leading to a severely limited sample.  In order to
be inclusive in such a search, a crucial condition for an accurate determination of the luminosity function,
a small degree of contamination can be tolerated provided the fraction is reasonably
well-understood.

In fact, the primary sources of contamination in high-redshift searches are now well-known.  We
briefly review them here.  Low-temperature galactic dwarf stars display quite similar colors to 
high redshift galaxies in the near infrared \citep[\eg,][]{bowler2012a}.  While ground-based 
surveys can only constrain stellar contamination by comparing SEDs of cool stars to 
observed colors \citep{bowler2012a},  WFC3 data has the distinct advantage of an extremely 
sharp point-spread function, ranging from 0.15$''$ in Y$_{105}$ to 0.18$''$ in H$_{160}$. Previous studies
in the UDF have utilized this to conclude that Galactic stars are not expected to contribute 
significantly ($\lt 2 \%$) to contamination in extremely deep, small area surveys 
\citep{finkelstein2010a,bouwens2011a}.

A more relevant concern is potential interlopers at $z \sim 2$.  Around
this redshift, the Balmer and 4000\AA\ breaks will lie near the same observed wavelengths where we
search for a Lyman break in our high-redshift sources.  However, unlike the Lyman break, the Balmer break
has a maximum possible depth, aiding us in isolating robust high-redshift sources (see Figure 2 of \citealt{kriek2010a} 
for actual measurements at $z > 0.5$).  Due to our extraordinarily deep optical data, these objects
must either then be severely reddened in the rest-frame ultraviolet or significantly affected by photometric
scatter in order to be picked up by our selection criteria.  We account for the possibility of such sources 
scattering into our selection window with our contamination simulations described below.

\subsection{Optical non-detection criteria}

Applying a rigorous set of optical non-detection criteria is key to obtaining a clean sample of high-redshift sources
by removing the lower redshift contaminants we describe above, as we expect all intrinsic flux from true high-redshift 
sources to be nearly extinguished by neutral Hydrogen at these wavelengths
\citep{madau1995a}.  Here we adopt a slightly modified version of the criteria used in \cite{bouwens2011a} to eliminate
sources with marginal optical detections from our selections.

The criteria we apply are as follows:  (1) The measured flux in each filter shortward of the dropout filter is less than 
2.0$\sigma$.  For the $z$-drops, this includes $B_{435}$, $V_{606}$, and $i_{775}$; for the $Y$-drops, we add $z_{850}$. 
(2)  No more than one of the filters listed above shows a detection above 1.5$\sigma$.  (3) To effectively
add all the optical data, we finally require $\chi^{2}_{opt}$\footnote{defined as $\chi^{2}_{opt} =$  $\Sigma_i sign(f_i) (f_i / \sigma_i)^2$, 
where the $i$ index runs across $B_{435}$, $V_{606}$, and $i_{775}$ for $z$-drops; 
additionally $i_{814}$ where available, and $z_{850}$ for $Y$-drops}
must be less than a threshold value, depending on the observed magnitude of the source.  We compute this
value using both our standard 0.30$''$ diameter circular apertures, and a smaller 0.21$''$ aperture, to 
rule out the most compact contaminants.  At or below the 5$\sigma$ limit of $m_{AB}$ = 29.5 (in uncorrected, aperture magnitudes), 
we adopt a $\chi^2_{opt}$ upper limit of 2.5, while at the  10$\sigma$ limit of $m_{AB}$ = 28.7, we relax this to limit 5.0.  
A linear interpolation is used to determine the limit for magnitudes between the 5 and 10$\sigma$ level. 

\subsection{Contamination simulations and the adopted UDF12 color-color selection}

To estimate the number of contaminants we expect, we utilized the excellent HST photometry for 
objects at 25.0 $< H_{160} < $ 27.0 (as measured in our $0.5''$ diameter apertures)
in the UDF.  As we can robustly rule out the presence of a Balmer break at this depth, we selected,
as our base color distribution of contaminants, all sources in this magnitude range that displayed at least
one optical detection.  The key assumption in our simulations is that the color distribution of these potential 
contaminants is unchanged as one moves down to the fainter magnitudes, where the majority of our dropout
galaxies lie.  Considering the trend in color in this magnitude range, we believe this may even be a conservative assumption.
For our sample at $25.00 < H_{160} < 25.25$, the median $z_{850} - Y_{105}$, $Y_{105} - J_{125}$, and $J_{125} - H_{160}$
colors are 0.45, 0.31, and 0.31, respectively.  At $26.75 < H_{160} < 27.00$, these three colors show even less extreme
median colors of 0.32, 0.17, and 0.15.

\begin{figure*}
\begin{center}
\includegraphics[width=0.45\textwidth]{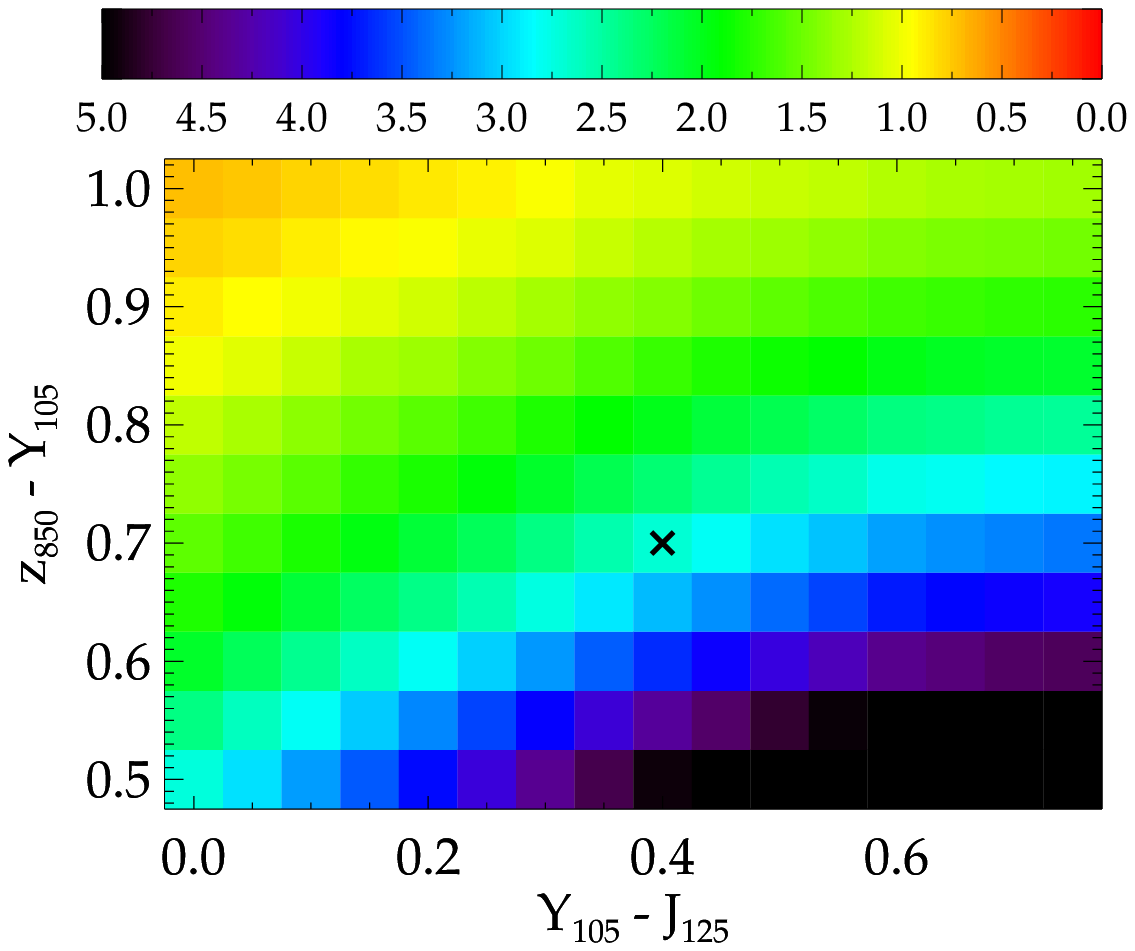}
\includegraphics[width=0.45\textwidth]{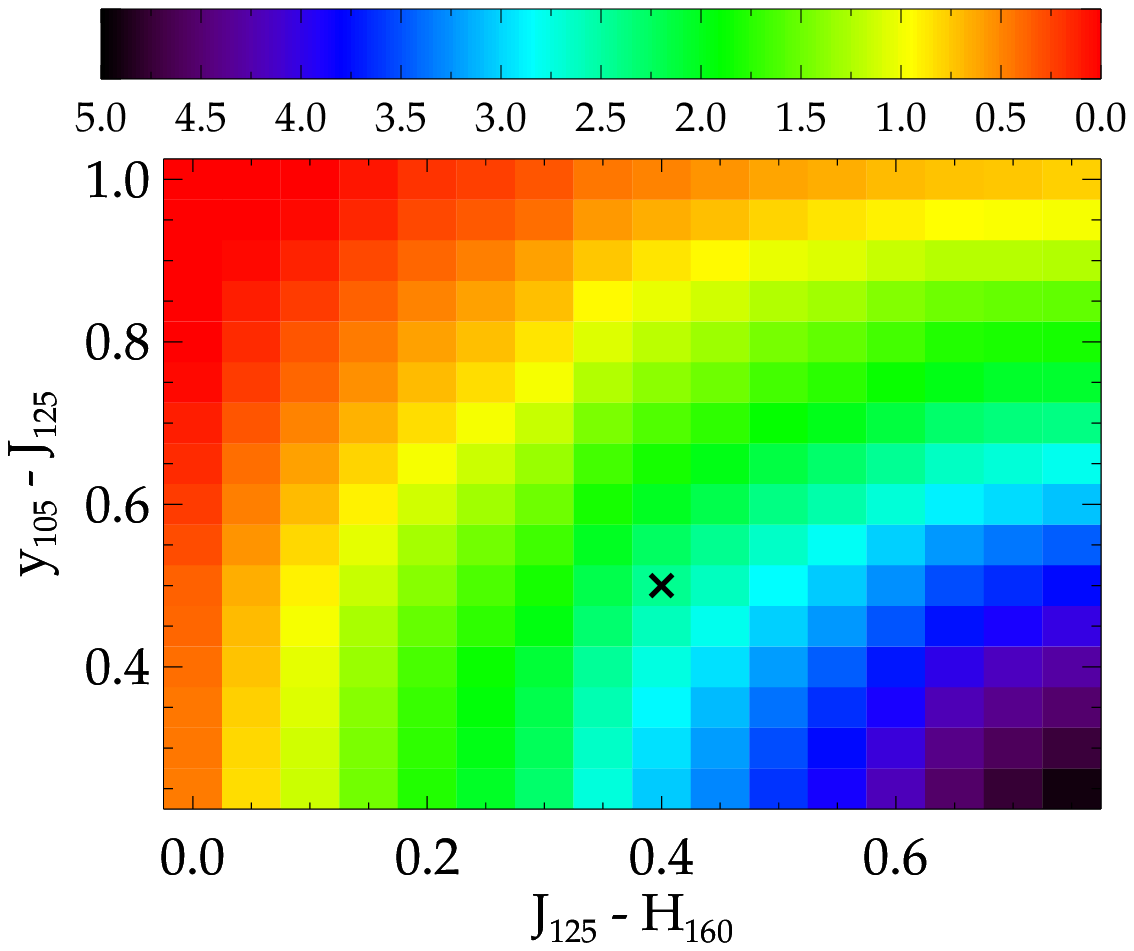}
\caption{\label{fig:col_cuts}Left: Number of $z_{850}$-drop contaminants per UDF12 field as a function of the color cuts in $z_{850}-Y_{105}$ and
$Y_{105}-J_{125}$.  The selection criteria are defined such that the $z_{850}-Y_{105}$ must be greater
than the value on the y-axis to be selected, and the$Y_{105}-J_{125}$ less than the value on the x-axis.  
Right: As left, but for $Y_{105}$-drops. Our chosen cuts are marked by the black $'$x$'$ in each figure.
We refer the reader to Figure \ref{fig:cand_plot} for a visualization of our final cuts.}
\end{center}
\end{figure*}

We then create an array of synthetic sources, matched to the actual number of observed sources in the 
UDF in bins of 0.1 magnitudes.  To get an accurate representation of sources intrinsically below our detection 
limit that have some chance of being scattered upwards into detection,
we extrapolate the number counts beyond $H_{160} = 28.7$ (equivalent to a source at $10.0\sigma$)  as a function of 
magnitude via a power law down to sources intrinsically as faint as $1.0\sigma$.  The colors of these
synthetic sources are chosen to obey the same color distribution as the brighter contaminant population 
described above, but noise consistent with their synthetic magnitudes is then added.  These sources
are then subject to the same optical non-detection criteria described above, but the colors cuts in 
$z_{850} - Y_{105}$ and $Y_{105} - J_{125}$ ($Y_{105}-J_{125}$ and $J_{125} - H_{160}$) for $z$-drops 
($Y$-drops) are varied in steps of 0.05 from as shown in Figure \ref{fig:col_cuts}.  To create these plots, we repeated 
the simulations for $N = 1000$ UDF fields.  

Clearly, the most robust constraint on eliminating contaminants comes from requiring a large break in the 
bluer color for each sample, but gains are also made by limiting the color longward of the break to relatively
blue values.  Provided with these estimates, we select chose to implement the color selection criteria
as shown in Figure \ref{fig:col_cuts}:

\begin{equation}
z_{850} - Y_{105} > 0.7
\end{equation}

\begin{equation}
Y_{105} - J_{125} < 0.4
\end{equation}

At $z \sim 8$, we use

\begin{equation}
Y_{105} - J_{125} > 0.5
\end{equation}

\begin{equation}
J_{125} - H_{160} < 0.4
\end{equation}

This leads to a selection for $z$-drops broadly within the redshift range $6.2<z<7.3$. For the
$z\simeq$8 study, we use our new ultra-deep WFC3 $Y_{105}$ as the dropout filter
which leads to a sample spanning the redshift range $7.3<z<8.5$. A discussion of more
distant sources in the UDF is provided by \citet{ellis2012a}.

Finally, to ensure our sources are robust, we demand a detection significance of 3.5$\sigma$ in the filter immediately
longward of the Lyman break ($Y_{105}$ for our $z$-drops, $J_{125}$ for our $Y$-drops, and a similarly
robust detection in one further WFC3 filter at longer wavelengths. 

These selection criteria are designed to provide as large a sample of galaxies as possible above redshift 6.5,
while minimizing the effect of contamination.  We plot selection functions for both $z$-drops and $Y$-drops
in Figure 2 (see Section 4.1 for details on the selection function simulation).  
At bright magnitudes $J_{140} < 27.5$, our $z$-drop color cuts provide a nearly complete 
census of star-forming galaxies between $6.30 < z < 7.15$, while the $Y$-drop cuts do the same
between $7.35 < z < 8.60$.  However, due to our strict optical non-detection criteria, and the area
subtended by other sources in UDF12, our maximum selection efficiency does not exceed $\sim 65 \%$.

Extending our $Y_{105}-J_{125}$ color cut for $z$-drops to 0.5 would only add
an extra $\sim$ 0.1 in redshift space to our selection function, as the color tracks are rapidly departing from our selection window
as can be seen in Figure 3.  This would additionally add an extra 0.3 expected contaminants.  Combined with the 
concern that our new $Y_{105}$ data is actually deeper than the existing $z_{850}$ data, and that the primary
contaminants themselves are intrinsically red, we opted for this conservative $Y_{105}-J_{125}$ color cut for our
$z$-drop sample.  Similarly adding an extra 0.1 mag to our $J_{125}-H_{160}$ limit for $Y$-drops 
is expected to add $\sim$ 0.4
contaminants.  As can be seen from the $Y$-drop color-color plot in Figure 3, the density of $Y$-drops
with $J_{125}-H_{160}$ colors this red is quite low, so we chose to truncate the selection at $J_{125}-H_{160} < 0.4$,
to limit the contamination.  Adopting these cuts and all of our previously discussed selection criteria, we
expect 2.79 and 1.42 contaminants per UDF field for $z$-drops and $Y$-drops, respectively.  

We include candidate lists and photometry for our final selection of $z$-drops and $Y$-drops in Tables 2 and 3,
respectively, and color-color plots in Figure 3.  In total, we select 47 $z$-drops, and 27 $Y$-drops, of which 
20 and 9 were only identified using our 
new UDF12 data.

\subsection{UDF-P1 and UDF-P2}

The two UDF parallel fields, observed as part of the UDF05 (GO10632; PI: Stiavelli) and UDF09 
(GO 11563; PI: Illingworth) campaigns comprise the two data sets most similar to our UDF12 dataset, though
the depths are $\sim$0.5 mag shallower.  As such, we utilize the same color-color criteria determined
for our UDF12 selections.  Because the optical data is shallower, we tighten our $\chi^2_{opt}$ upper limit
for selection to 1.5 (3.0) at the 5$\sigma$ (10$\sigma$) aperture magnitude limit in each field.  We 
utilize the ultradeep 128 orbit $z_{814}$ ACS data taken in parallel to our main UDF12 program, which covers roughly 70 $\%$
of the P2 WFC3/IR field.  For $z$-drops within this area we impose an additional criteria of $z_{814} - Y_{105} > 2.0$
OR $f_{814}/\sigma_{814} < 2.0$.

\subsection{ERS}

For the ERS dataset, the $Y_{098}$ filter was utilized, so we take care to alter our 
selection criteria accordingly.  We chose to use the criteria derived by \cite{bouwens2011a}, at $z \sim 7$, which,
despite the different filter selection, produce a selection function that probes a similar range in redshift.  These criteria
are (1) $z_{850} - J_{125} > 0.9$, (2) $z_{850} - J_{125} > 0.8 + 1.1(J_{125}-H_{160})$, and (3) $z_{850}-J_{125} > 0.4 
+ 1.1(Y_{098}-J_{125})$.  At $z \sim 8$, we adopt the \cite{bouwens2011a}  $Y_{098} - J_{125}$ lower limit of 1.25, but 
chose the same $J_{125} - H_{160} < 0.4$ cut we use for the UDF fields, to ensure the selection functions for our analysis
are as homogeneous as possible.

\subsection{CANDELS}

For the CANDELS field, we use the same criteria as adopted for UDF-P34, including the addition of the $z_{814}$
criteria.  Due to the varying depth of the data, we divide the CANDELS field into 3 distinct subregions when performing
our analysis.  The first region consists of the immediate area around the HUDF that is covered by ACS optical imaging
taken as part of the HUDF04 \citep{beckwith2006a} campaign, but falls outside the region probed by our UDF12
campaign, due to the smaller field of view of the WFC3 instrument.  Here, the optical data is nearly a full 2 magnitudes
deeper than the available IR data, resulting in negligible contamination rates in our sample.  

The remaining sample is further divided into an East and West region.  The East and West regions have identical depth
in the ACS data from the GOODS program, and in the $J_{125}$ and $H_{160}$ filters, but the $Y_{105}$ depth in the
West region is approximately 0.4 magnitudes shallower than that of the East.  For the purposes of contamination and
completeness simulations, we separate these two fields in order to properly account for the variation in depth.

\section{The Luminosity Function at $z \sim 7$ and $\sim$ 8 from UDF12 data}

With our candidate selection completed, we now turn to evaluating the $z\simeq$ 7 and 8 luminosity
functions. The key issue in converting numbers of sources of known absolute magnitude into a comoving
density of luminosities is, of course, the redshift-dependent selection function which defines the visibility
as a function of apparent magnitude as well as the optimum algorithm for fitting a function to the resulting
number density. We now introduce the algorithms we will utilize for both of these critical steps.

\subsection{Simulations}
\label{section:simulations}

We first describe how we calculate the selection function used to determine the effective volumes for our 
luminosity function calculations.

To create synthetic fluxes for galaxies in our simulations, we assume an input UV slope $\beta = -2.0$.  This 
choice is motivated by the results of \cite{dunlop2012b}, who find no 
conclusive evidence for an intrinsic scatter in $\beta$ from this
value at $z\sim 7-8$.  We compute fluxes applying intergalactic extinction from \cite{meiksen2006a} to a 
Bruzual and Charlot \citep{bruzual2003a} synthetic spectrum consistent with this value of $\beta$.

We parameterize our selection efficiency, $S(m,z)$, as a function of redshift and the magnitude for the
filter (or filters for the UDF12 data) used to determine the rest-frame UV magnitude.  
 In each field, we determine the shape of the selection function first using numerical
simulations only, which take into account the limiting magnitudes in each filter for point sources.  In steps of 
0.01 in redshift, and 
0.05 in magnitude, we take the synthetic flux from our galaxy model, perturb it by the appropriate noise, and 
apply our color-color selection criteria.  At each step of $z,m$, and for each field, we perform this $N=1000$ 
times to construct a complete surface for our selection function.  We define the selection function produced by
this process as $S_{numeric}(m,z)$.

However, this selection function is only appropriate if these galaxies are point sources in otherwise blank fields, 
which is certainly not the case.  The marginally resolved nature of our targets will result in higher incompleteness
levels at faint magnitudes than for point sources.  It is imperative to correct for this effect, as a varying
completeness correction can produce large differences in the faint end slope \citep{grazian2012a}.  To
account for this incompleteness, as well as that caused by area in the images obscured by brighter sources,
we rely on inserting synthetic galaxy images into our mosaics for each field.  We generate a synthetic
template with a Sersic index of 1.5, consistent with a stack of LBGs at $z\sim4$ \citep{oesch2010a}. 
The template image has a half-light radius of 0.35 kpc, motivated by the results of \cite{ono2012a} 
who perform a detailed measurement
of the sizes of UDF12 high-redshift candidates.  This template is then convolved with the point spread function
for each filter, multiplied by the appropriate model flux as described above, and inserted into the image in a 
random location.  After inserting $N=1000$ non-overlapping sources, we run the SExtractor program for object
detection, and compute fluxes and errors in the same manner as for our science images.  We perform this 
simulation at the peak of each of our selection functions in redshift space, in steps of 0.05 in magnitude.  This peak
efficiency at a given magnitude $\epsilon(m)$ is then used to normalize our selection function such that our
total selection function used to compute effective volumes is given by
$S_{total}(m,z) = S_{numeric}(m,z) \times \epsilon(m)$.  The final selection functions for both $z$-zdrops and 
$Y$-drops in the UDF are presented in Figure 2.

\subsection{Maximum Likelihood Luminosity Functions}

Using the new UDF12 data and previous observations,
we assemble dropout samples at $z \sim 7$ and $\sim 8$ in multiple fields.
For each sample, we split the galaxy number counts into bins of width 
$\Delta M = 0.5$ in absolute magnitude M$_{UV}$.  
We use these binned galaxy number counts and our simulations of photometric
scattering and low-redshift contaminants to determine the high-redshift
stepwise maximum likelihood (SWML; \citealt{efstathiou1988a}) and \cite{schechter1976a}
galaxy luminosity functions $\Phi(\MUV)$ in units of $\phiunit$.

\subsubsection{Stepwise Maximum Likelihood Luminosity Function}
\label{section:swml}

The SWML luminosity function aims to jointly fit the binned galaxy
abundance $\Phik$ in the $k^{\mathrm{th}}$ of $\Nbin$ magnitude
bins. For each bin, the maximum likelihood values for $\Phik$ are determined by
using the observed number of galaxies $\nobsk$, the effective volume
$\Veffk$ for galaxies with intrinsic $\MUV$ in the bin, and the 
probabilities $P_{i,j}$ for scattering galaxies with intrinsic $\MUV$ in bin $i$ 
into bin $j$
owing to photometric noise, and the number of low-redshift contaminants
$\nconk$ for the bin calculated as described in \S \
{section:simulations}.
Given $\Phik$, we construct the expected 
number of galaxies in each bin as
\begin{eqnarray}
\label{eqn:nexpk}
\nexpk &=& \Phik \Veffk\left(1 - \sum_{i\ne k} P_{k,i}\right)\nonumber\\
&&+ \sum_{i \ne k } \Phi_i V_{\mathrm{eff},i} P_{i,k} \nonumber\\
&&+ \nconk, 
\end{eqnarray}
\noindent
where the summations run over $\Nbin$.
In practice, when using a bin width $\Delta \MUV = 0.5$ galaxies do not scatter
by more than one bin and the summations are trivial. We account for photometric
scattering of sources into our faintest bin by a simple extrapolation of the
luminosity function to yet fainter magnitudes.

To fit the shape of our SWML, we can use the likelihood of observing $\nobsk$ given the 
expected number $\nexpk$:

\begin{equation}
\label{eqn:poisson}
p(\nobsk | \nexpk) = \left(\frac{\nexpk}{\Sigma_j \nexpj}\right)^{\nobsk}
\end{equation}
\noindent

The SWML luminosity function $\Phik$ is determined by maximizing the
product of these individual likelihoods across all bins, and across
all fields. Photometric scatter correlates
the individual $\Phik$ for each field, and the SWML values must therefore be determined 
simultaneously.  This calculation amounts to an 
$\Nbin$-parameter estimation problem, and we use the {\it MultiNest} 
nested sampling code for Bayesian inference problems \citep{feroz2008a, feroz2009a}
to maximize the $\Phik$ likelihoods.

To fit the overall normalization of our SWML, we sum $\nobsk$ across all fields for each
magnitude bin.  Since we expect this quantity to be Poisson distributed, we can easily generate
a posterior distribution of the normalization for each bin.  To find our final posterior distribution, 
for the normalization, we multiply the posteriors generated in this manner across all bins.

The results of the SWML luminosity function calculation for redshift $z\sim7$ 
is shown in Figure \ref{fig:z7_lf} and for redshift $z\sim8$ in Figure \ref{fig:z8_lf}.
The data points indicate the maximum likelihood $\Phik$ at each magnitude for 
our sample taking into account all fields, while error bars indicate the smallest
marginalized interval to encompass $68\%$ of the likelihood for each bin.

\subsubsection{Schechter Luminosity Functions}
\label{section:schechter}

A determination of the \cite{schechter1976a} luminosity function parameters
is calculated by estimating the mean galaxy abundance in each bin as
\begin{equation}
\Phik = \frac{1}{\Delta \MUV} \int_{M_{\mathrm{UV},k}-0.5\Delta\MUV}^{M_{\mathrm{UV},k}+0.5\Delta\MUV} \Phi(M) dM
\end{equation}
\noindent
and then using Equations \ref{eqn:nexpk} and \ref{eqn:poisson} to calculate
the likelihood of the fit parameters. The likelihoods of each binned sample
in each field are multiplied.
To improve constraints at the bright end at $z \sim 7$, when fitting a we also include data from the ground-based
surveys of \cite{ouchi2009a}, \cite{castellano2010a} and \cite{bowler2012a}.  Incorporating the wide-area ground-based
constraints
is critical as even our wide area 
HST data only detects sources dimmer than $M_{\mathrm{UV}} \sim -21.0$, or approximately 1 magnitude brighter than
$M^{*}_{UV}$ at this redshift.  As pointed out in \cite{robertson2010b}, \cite{bouwens2011a}, and \cite{bradley2012a}, there remains a large
degeneracy between $M^{*}_{UV}$ and the faint end slope, $\alpha$ without sufficient data at the bright end.  
These additional data are incorporated using the published data points and errors through a $\chi^{2}$ likelihood, assuming the
reported errors are Gaussian. The maximum likelihood Schechter function parameters are determined using 
{\it MultiNest} to conduct Bayesian inference.

The Schechter function fit results for redshift $z\sim7$ are shown 
in Figure \ref{fig:z7_lf} and for $z\sim8$ in 
Figure \ref{fig:z8_lf}, with the maximum likelihood models shown as a black line.
At $z\sim7$, the best fit Schechter function parameters are $\log_{10} \phi_{\star} = -3.19^{+0.27}_{-0.24} \log_{10} \phiunit$,
$M_{\mathrm{UV},\star} = -20.1 \pm 0.4 $, and $\alpha_{z\sim7} = -1.87^{+0.18}_{-0.17} $.  The uncertainty range for
each parameter reflects the smallest interval in each marginalized distribution to encompass 68\% of the posterior probability.  
At $z\sim8$, the best fit Schechter function parameters are $\log_{10} \phi_{\star} = -3.50^{+0.35}_{-0.32} \log_{10} \phiunit$,
$M_{\mathrm{UV},\star} = -20.4^{+0.5}_{-0.4}  $, and $\alpha_{z\sim8} = -1.94^{+0.21}_{-0.24}$.
In Figures \ref{fig:z7_lf} and \ref{fig:z8_lf}, the grey shaded regions denote the inner 68$\%$ of the marginalized
posterior distribution in galaxy abundance at each magnitude.  

We caution the reader against an 
over-interpretation of the faintest bins in our Luminosity functions.  Although heating of the intergalactic
medium during reionization is expected to suppress the formation of dwarf galaxies below a characteristic
halo mass \citep[e.g.,][]{wyithe2006a,munoz2011a}, the density determinations of $\phi_k$ in our faintest
bins are very sensitive both to upscattering of sources below the limit and completeness corrections.  This is
largely a result of being in a regime where the effective volume is rapidly changing as a function of magnitude.  
For example, simply
dividing the number of observed sources, after correcting for the expected contamination, in our faintest $z \sim 7$ 
bin by the effective volume yields $\log_{10} \phi_k \sim -2.3$, much more in line with our best fit Schechter
function parameters.  Though we have made our best effort to quantify and account for corrections arising from
finite size, scattering, and contamination, the situation remains difficult at the faintest reaches of the survey.   
 
\subsubsection{Cosmic Variance}
\label{section:cv}

Although we have not included the effects of cosmic variance in any of the parameter estimates given above,
it nonetheless useful to obtain some indication of its effect.  To first order, cosmic variance is unlikely to have
a major effect on one of the primary goals of this paper, namely an estimate of the faint end slope at $z \sim7-8$.
In the following, we therefore give an approximate 
calculation of the effective variance arising from large scale structure. 

Density fluctuations owing to large scale modes can cause variations in the observed galaxy abundance beyond
uncertainties arising from number counting statistics.  Following \cite{robertson2010b}, by using our best
fit luminosity functions at $z\sim7-8$, we can characterize
these cosmic variance uncertainties for each field in our sample.  We use the linear power spectrum calculated
with the \cite{eisenstein1998a} transfer function, conservatively assuming root-mean-squared density fluctuations
in volumes of radius $8~h^{-1}\mathrm{Mpc}$ of $\sigma_{8}=0.9$ at $z=0$, to estimate the typical root-mean-squared 
density fluctuations in our survey fields at the redshifts of interest. To estimate the clustering bias of galaxies
at these redshifts, we simply match the abundance of galaxies
from our luminosity functions with the abudance of dark matter halos provided by the \citet{tinker2008a}
halo mass function, and then assign the clustering bias of the halos to the galaxies assuming the bias
function of \cite{tinker2010a}.  
For the UDF, to our limiting magnitude
we find that the typical cosmic variance (the fractional uncertainty in the total galaxy number
counts owing to large scale structure) is $\sigma_{CV}\approx0.30$
at $z\sim7$ and $\sigma_{CV}\approx0.36$ at $z\sim8$.  The typical bias for galaxies in the UDF
is $b\approx5.0$ at $z\sim7$ and $b\approx 6.2$ at $z\sim8$.
For UDF-P1, 
we find that the typical cosmic variance is $\sigma_{CV}\approx0.33$
at $z\sim7$ and $\sigma_{CV}\approx0.38$ at $z\sim8$.  The typical bias for galaxies in the UDF-P1
is $b\approx5.4$ at $z\sim7$ and $b\approx 6.6$ at $z\sim8$.
For UDF-P2, 
we find that the typical cosmic variance is $\sigma_{CV}\approx0.32$
at $z\sim7$ and $\sigma_{CV}\approx0.37$ at $z\sim8$.  The typical bias for galaxies in the UDF-P2
is $b\approx5.2$ at $z\sim7$ and $b\approx 6.5$ at $z\sim8$.
For ERS,
we find that the typical cosmic variance is $\sigma_{CV}\approx0.30$
at $z\sim7$ and $\sigma_{CV}\approx0.34$ at $z\sim8$.  The typical bias for galaxies in the ERS field
is $b\approx6.4$ at $z\sim7$ and $b\approx 7.7$ at $z\sim8$.
Lastly, for CANDELS-Deep 
we find that the typical cosmic variance is $\sigma_{CV}\approx0.26$
at $z\sim7$ and $\sigma_{CV}\approx0.30$ at $z\sim8$.  The typical bias for galaxies in CANDELS-Deep
is $b\approx6.3$ at $z\sim7$ and $b\approx 7.6$ at $z\sim8$.

\section{Discussion}

It is instructive to compare the Schechter function parameters derived by our study to those of previous analyses,
both at these redshifts, and below.  We focus here on the faint end slope, $\alpha$, as this is the primary
parameter for which we expect our new UDF12 program to provide an improved constraint.  Previous studies
at redshifts $2 < z < 6$ find a remarkably consistent value of $\alpha \sim -1.7$ across this range 
\citep[\eg,][]{oesch2007a, bouwens2007a, reddy2009a, mclure2009a}.  

At the moderately larger redshift of
$z \sim 7$, the situation remains much more uncertain.  While \cite{bouwens2011a} claim a significantly 
steepened value of $\alpha = -2.01 \pm 0.21$, at $z \sim 7$, \cite{grazian2012a} find no signal of slope evolution, 
determining $\alpha = -1.7 \pm 0.1$.  Our determination of $\alpha_{z\sim7} = -1.87^{+0.18}_{-0.17} $, though 
still consistent with $-1.7$, does suggest a 
steepening of the faint end slope with increasing redshift, especially when considering the value of $\alpha =-1.90^{+0.14}_{-0.15}$ determined by
\cite{mclure2012a}, who also use the new UDF12 data, but perform a completely independent analysis.  
It is reassuring that the two methods agree so closely, despite the differing selection techniques and 
different potential sources of uncertainty. 

At $z \sim 8$, the existing literature largely agrees on a steepening of $\alpha$, with the most recent determinations
by \cite{bouwens2011a}, \cite{oesch2012a} and \cite{bradley2012a} finding values between $-1.9$ and $-2.1$. 
Extending 0.25 magnitudes fainter in UV luminosity than any previous study, our determination of 
$\alpha_{z\sim8} =-1.94^{+0.21}_{-0.24}$ provides increased support for this evolution, in concert with the
$\alpha_{z\sim8} = -2.02^{+0.22}_{-0.23}$ found by \cite{mclure2012a}. As noted by many 
authors\citep[\eg,][]{robertson2010a,bouwens2012a}, this
will significantly increase the ability of galaxies to maintain the reionization of the
intergalactic medium as intrinsically faint sources become more numerous.  This steepening
is also predicted by conditional luminosity function methods based on the evolution of the dark matter
halo mass function \citep{trenti2010a}.  We also note that although our derived values of 
$\phi_{\star}$ and $M_{\mathrm{UV},\star} $ favor an decreasing $\phi_{\star}$ with
redshift to account for the evolution of the luminosity function, the errors are still too large
to rule out an evolution in the characteristic magnitude instead (or a combination).
We defer a significantly more detailed analysis of the impact of our survey on reionization
to \cite{robertson2012a}.

Along with \cite{mclure2012a}, we have uncovered the most comprehensive and robust sample of subluminous
high-redshift galaxies to date.  At moderate magnitudes, $M_{\mathrm{UV}} \leq -18.0$, we achieve a more refined
sample of dropouts, including an additional 3 (3) $z$-drops ($Y$-drops) not previously identified as high redshift sources
as a result of our improved photometry.  Of greater importance, though, are our advances below this UV magnitude; we
discover an additional 14 sources at $z \sim 7$ by virtue of our ultradeep $Y_{105}$ image, as well as an additional
5 sources at $z \sim 8$, indicating the steepness of the faint end slope continues beyond 2 magnitudes below 
$M_{\mathrm{UV},\star}$ at these redshifts.  Additionally, our sample is in excellent agreement with the independent
determination from \cite{mclure2012a}.  We note only 2 of our sources at $z\sim7$, and 3 at $z\sim8$ are not
present in their final catalog.

With the upcoming HST \textit{Frontier Fields} observations scheduled to begin in Cycle 21, progress in this
regime vital to understanding if and when starforming galaxies can maintain reionization is sure to continue.
We stress the gains made by UDF12 strengthen claims of an increased steepness at the faint end and, along
with \cite{mclure2012a} provide a self-consistent, robust determination of $\alpha$ at redshifts 7 and 8.

\acknowledgements

US authors acknowledge financial support from the Space Telescope Science Institute under award HST-GO-12498.01-A. 
JSD acknowledges support of the European Research Council and the Royal Society. RJM acknowledges funding
from the Leverhulme Trust. SC acknowledges the support of the European Commission through the Marie Curie Initial Training Network ELIXIR.  This work is based on data from the {\it Hubble Space Telescope} 
operated by NASA through the Space Telescope Science Institute via the association of
Universities for Research in Astronomy, Inc.\@ under contract NAS5-26555.

\begin{deluxetable*}{lccccccc}
\tablecolumns{8}
\tablewidth{0pt}
\tablecaption{\bf Datasets}
\tablehead{\colhead{Name} & \colhead{Area (arcmin$^2$)} & \colhead{z$\sim$7 candidates}
& \colhead{z$\sim$8 candidates} & \colhead{$Y_{105W}$} & \colhead{$J_{125W}$} 
& \colhead{$J_{140W}$} & \colhead{$H_{160W}$}}
\medskip
\startdata
\multicolumn{8}{r}{\bf Survey Depth 5-{\bf $ \sigma$} AB}\medskip \\
UDF12    & 4.6  & 47    & 27  & 30.0  & 29.5  & 29.5  & 29.5    \\
UDF-P1   & 4.6   & 15    & 10  & 28.9  & 29.0  &  -    & 28.7    \\
UDF-P2   & 4.6   & 13    & 10  & 29.0  & 29.2  &  -    & 28.7    \\
ERS        & 36.8  & 15    & 4   & 27.6\tablenotemark{a}  & 28.0  &  -    & 27.5    \\
CANDELS-Deep   & 64.9  & 32    & 18  & 28.2  & 28.1  &  -    & 27.7    \\
\smallskip
\enddata
\tablecaption{}
\tablenotetext{a} {The ERS and program replaces the $Y_{105}$ filter with the $Y_{098}$ filter.}
\end{deluxetable*}

\begin{deluxetable*}{lcccccccc}
\tablecolumns{9}
\tablewidth{0pt}
\tablecaption{\bf $z$-drops $6.2<z<7.3$}
\tablehead{\colhead{ID} & \colhead{RA} & \colhead{Dec}
& \colhead{$z_{850LP}$} & \colhead{$Y_{105W}$} & \colhead{$J_{125W}$} 
& \colhead{$J_{140W}$} & \colhead{$H_{160W}$} & \colhead{References}}
\medskip
\startdata

UDF12-3746-6328	 & 3:32:37.46	 & -27:46:32.8	 & 28.4 $\pm$ 0.1	 & 27.6 $\pm$ 0.1\tablenotemark{a}	 & 27.5 $\pm$ 0.1\tablenotemark{a}	 & 27.4 $\pm$ 0.1\tablenotemark{a}	 & 27.3 $\pm$ 0.1\tablenotemark{a}	 & 1,3,9  \\
UDF12-4256-7314	 & 3:32:42.56	 & -27:47:31.4	 & 30.4 $\pm$ 0.7	 & 28.1 $\pm$ 0.1\tablenotemark{a}	 & 27.8 $\pm$ 0.1\tablenotemark{a}	 & 27.8 $\pm$ 0.1\tablenotemark{a}	 & 27.7 $\pm$ 0.1\tablenotemark{a}	 & 1,2,3,4,5,7,10  \\
UDF12-4219-6278	 & 3:32:42.19	 & -27:46:27.8	 & 29.2 $\pm$ 0.2	 & 28.1 $\pm$ 0.1\tablenotemark{a}	 & 28.1 $\pm$ 0.1	 & 28.0 $\pm$ 0.1	 & 28.1 $\pm$ 0.1\tablenotemark{a}	 & 1,3,4,7,9,10  \\
UDF12-3677-7536	 & 3:32:36.77	 & -27:47:53.6	 & 29.0 $\pm$ 0.2	 & 28.2 $\pm$ 0.1\tablenotemark{a}	 & 28.2 $\pm$ 0.1	 & 28.3 $\pm$ 0.1	 & 28.2 $\pm$ 0.1\tablenotemark{a}	 & 1,2,3,4,7,9,10  \\
UDF12-3744-6513	 & 3:32:37.44	 & -27:46:51.3	 & 29.5 $\pm$ 0.2	 & 28.4 $\pm$ 0.1\tablenotemark{a}	 & 28.3 $\pm$ 0.1	 & 28.5 $\pm$ 0.1	 & 28.5 $\pm$ 0.1	 & 1,2,3,4,5,7,9,10  \\
UDF12-4105-7156	 & 3:32:41.05	 & -27:47:15.6	 & 30.4 $\pm$ 0.7	 & 28.7 $\pm$ 0.1	 & 28.4 $\pm$ 0.1	 & 28.3 $\pm$ 0.1	 & 28.4 $\pm$ 0.1	 & 1,2,3,5,10  \\
UDF12-3958-6565	 & 3:32:39.58	 & -27:46:56.5	 & 29.8 $\pm$ 0.3	 & 28.4 $\pm$ 0.1\tablenotemark{a}	 & 28.4 $\pm$ 0.1	 & 28.4 $\pm$ 0.1	 & 28.4 $\pm$ 0.1	 & 1,2,3,5,7,9,10  \\
UDF12-3638-7162	 & 3:32:36.38	 & -27:47:16.2	 & 29.5 $\pm$ 0.2	 & 28.5 $\pm$ 0.1\tablenotemark{a}	 & 28.5 $\pm$ 0.1	 & 28.4 $\pm$ 0.1	 & 28.5 $\pm$ 0.1	 & 1,2,3,4,5,7,9,10  \\
UDF12-4057-6436	 & 3:32:40.57	 & -27:46:43.6	 & 29.9 $\pm$ 0.4	 & 28.6 $\pm$ 0.1	 & 28.6 $\pm$ 0.1	 & 28.5 $\pm$ 0.1	 & 28.7 $\pm$ 0.1	 & 1,2,3,4,5,7,9,10  \\
UDF12-4431-6452	 & 3:32:44.31	 & -27:46:45.2	 & 29.6 $\pm$ 0.3	 & 28.7 $\pm$ 0.1	 & 28.6 $\pm$ 0.1	 & 28.7 $\pm$ 0.1	 & 28.9 $\pm$ 0.2	 & 1,7,9,10  \\
UDF12-4160-7045	 & 3:32:41.60	 & -27:47:04.5	 & 30.0 $\pm$ 0.5	 & 28.8 $\pm$ 0.1	 & 28.8 $\pm$ 0.1	 & 28.7 $\pm$ 0.1	 & 28.8 $\pm$ 0.1	 & 7,9,10  \\
UDF12-4268-7073	 & 3:32:42.68	 & -27:47:07.3	 & 29.9 $\pm$ 0.3	 & 29.1 $\pm$ 0.1	 & 28.9 $\pm$ 0.1	 & 28.9 $\pm$ 0.2	 & 28.9 $\pm$ 0.1	 & 10  \\
UDF12-3313-6545	 & 3:32:33.13	 & -27:46:54.5	 &  $<$ 31.3	 & 29.3 $\pm$ 0.1	 & 28.9 $\pm$ 0.1	 & 28.6 $\pm$ 0.1	 & 29.0 $\pm$ 0.1	 & 1,6,7,9,10  \\
UDF12-3402-6504	 & 3:32:34.02	 & -27:46:50.4	 & 30.8 $\pm$ 0.7	 & 29.3 $\pm$ 0.1	 & 28.9 $\pm$ 0.1	 & 29.0 $\pm$ 0.1	 & 29.2 $\pm$ 0.2	 & 7,10  \\
UDF12-4182-6112	 & 3:32:41.82	 & -27:46:11.2	 & 30.3 $\pm$ 0.5	 & 29.3 $\pm$ 0.1	 & 29.1 $\pm$ 0.1	 & 29.0 $\pm$ 0.1	 & 28.9 $\pm$ 0.1	 & 3,4,5,7,9,10  \\
UDF12-3734-7192	 & 3:32:37.34	 & -27:47:19.2	 & 30.3 $\pm$ 0.5	 & 29.3 $\pm$ 0.1	 & 29.2 $\pm$ 0.2	 & 29.5 $\pm$ 0.2	 & 29.5 $\pm$ 0.2	 & 10  \\
UDF12-4239-6243	 & 3:32:42.39	 & -27:46:24.3	 & 30.0 $\pm$ 0.3	 & 29.2 $\pm$ 0.1	 & 29.3 $\pm$ 0.2	 & 28.9 $\pm$ 0.1	 & 29.1 $\pm$ 0.2	 & 3,5,7,11  \\
UDF12-3989-6189	 & 3:32:39.89	 & -27:46:18.9	 & 30.5 $\pm$ 0.8	 & 29.4 $\pm$ 0.1	 & 29.3 $\pm$ 0.2	 & 29.5 $\pm$ 0.2	 & 29.5 $\pm$ 0.2	 & 1,10  \\
UDF12-4068-6498	 & 3:32:40.68	 & -27:46:49.8	 & 30.6 $\pm$ 0.8	 & 29.7 $\pm$ 0.1	 & 29.4 $\pm$ 0.2	 & 29.4 $\pm$ 0.2	 & 29.6 $\pm$ 0.2	 & 10  \\
UDF12-3853-7519	 & 3:32:38.53	 & -27:47:51.9	 &  $<$ 31.4	 & 29.6 $\pm$ 0.1	 & 29.4 $\pm$ 0.2	 & 29.4 $\pm$ 0.2	 & 29.7 $\pm$ 0.3	 & 2,7,10  \\
UDF12-4472-6362	 & 3:32:44.72	 & -27:46:36.2	 & 30.8 $\pm$ 1.0	 & 29.0 $\pm$ 0.1	 & 29.4 $\pm$ 0.3	 & 28.8 $\pm$ 0.1	 & 28.9 $\pm$ 0.1	 & 10  \\
UDF12-3983-6189	 & 3:32:39.83	 & -27:46:18.9	 & 30.0 $\pm$ 0.4	 & 29.2 $\pm$ 0.1	 & 29.4 $\pm$ 0.2	 & 29.7 $\pm$ 0.3	 & 29.4 $\pm$ 0.1	 & 1,11  \\
UDF12-3736-6245	 & 3:32:37.36	 & -27:46:24.5	 & 30.6 $\pm$ 0.7	 & 29.4 $\pm$ 0.1	 & 29.5 $\pm$ 0.2	 & 29.5 $\pm$ 0.2	 & 29.6 $\pm$ 0.2	 & 7,10  \\
UDF12-3456-6493	 & 3:32:34.56	 & -27:46:49.3	 & 31.4 $\pm$ 3.9	 & 29.6 $\pm$ 0.1	 & 29.5 $\pm$ 0.2	 & 29.3 $\pm$ 0.1	 & 29.2 $\pm$ 0.1	 & 7,10  \\
UDF12-3859-6521	 & 3:32:38.59	 & -27:46:52.1	 & 30.4 $\pm$ 0.6	 & 29.3 $\pm$ 0.1	 & 29.5 $\pm$ 0.2	 & 29.9 $\pm$ 0.3	 & 29.5 $\pm$ 0.2	 & 7,10  \\
UDF12-4384-6311	 & 3:32:43.84	 & -27:46:31.1	 &  $<$ 30.9	 & 29.8 $\pm$ 0.2	 & 29.6 $\pm$ 0.3	 & 29.5 $\pm$ 0.2	 & 29.9 $\pm$ 0.4	 & 10  \\
UDF12-3755-6019	 & 3:32:37.55	 & -27:46:01.9	 &  $<$ 31.4	 & 29.9 $\pm$ 0.2	 & 29.6 $\pm$ 0.2	 & 29.9 $\pm$ 0.4	 & 29.6 $\pm$ 0.2	 & 7,10  \\
UDF12-3975-7451	 & 3:32:39.75	 & -27:47:45.1	 & 30.9 $\pm$ 1.1	 & 29.4 $\pm$ 0.1	 & 29.6 $\pm$ 0.2	 & 29.2 $\pm$ 0.2	 & 29.3 $\pm$ 0.2	 & 7,10  \\
UDF12-4201-7074	 & 3:32:42.01	 & -27:47:07.4	 & 30.6 $\pm$ 1.0	 & 29.6 $\pm$ 0.1	 & 29.6 $\pm$ 0.2	 & 30.0 $\pm$ 0.4	 & 29.6 $\pm$ 0.2	 & 10  \\
UDF12-4037-6560	 & 3:32:40.37	 & -27:46:56.0	 &  $<$ 31.3	 & 29.9 $\pm$ 0.2	 & 29.6 $\pm$ 0.2	 & 30.1 $\pm$ 0.4	 & 30.1 $\pm$ 0.3	 & 7,10  \\
UDF12-4426-6367	 & 3:32:44.26	 & -27:46:36.7	 & 30.8 $\pm$ 1.4	 & 29.6 $\pm$ 0.1	 & 29.6 $\pm$ 0.3	 & 29.7 $\pm$ 0.3	 & 29.5 $\pm$ 0.2	 & -  \\
UDF12-3909-6092	 & 3:32:39.09	 & -27:46:09.2	 & 30.2 $\pm$ 0.6	 & 29.5 $\pm$ 0.1	 & 29.7 $\pm$ 0.2	 & 30.3 $\pm$ 0.4	 & 29.9 $\pm$ 0.3	 & 10  \\
UDF12-4143-7041	 & 3:32:41.43	 & -27:47:04.1	 & 30.6 $\pm$ 0.7	 & 29.9 $\pm$ 0.2	 & 29.8 $\pm$ 0.2	 & 30.2 $\pm$ 0.4	 & 30.0 $\pm$ 0.4	 & 10  \\
UDF12-3696-5536	 & 3:32:36.96	 & -27:45:53.6	 & 30.5 $\pm$ 0.5	 & 29.7 $\pm$ 0.2	 & 29.8 $\pm$ 0.3	 & 29.3 $\pm$ 0.2	 & 29.9 $\pm$ 0.3	 & 7,10  \\
UDF12-3897-8116	 & 3:32:38.97	 & -27:48:11.6	 & 30.7 $\pm$ 0.8	 & 29.9 $\pm$ 0.2	 & 29.8 $\pm$ 0.3	 & 30.0 $\pm$ 0.4	 & 29.8 $\pm$ 0.3	 & 10  \\
UDF12-4288-6261	 & 3:32:42.88	 & -27:46:26.1	 & 30.7 $\pm$ 1.0	 & 29.7 $\pm$ 0.2	 & 29.8 $\pm$ 0.2	 & 30.0 $\pm$ 0.3	 & 30.5 $\pm$ 0.7	 & 10  \\
UDF12-3817-7327	 & 3:32:38.17	 & -27:47:32.7	 &  $<$ 30.5	 & 30.0 $\pm$ 0.2	 & 29.8 $\pm$ 0.3	 & 30.0 $\pm$ 0.4	 & 30.3 $\pm$ 0.6	 & 11  \\
UDF12-4379-6510	 & 3:32:43.79	 & -27:46:51.0	 & 30.8 $\pm$ 1.2	 & 29.4 $\pm$ 0.1	 & 29.9 $\pm$ 0.3	 & 29.8 $\pm$ 0.3	 & 29.8 $\pm$ 0.3	 & 10  \\
UDF12-3691-6517	 & 3:32:36.91	 & -27:46:51.7	 &  $<$ 30.7	 & 30.0 $\pm$ 0.2	 & 29.9 $\pm$ 0.3	 & 29.7 $\pm$ 0.3	 & 30.2 $\pm$ 0.4	 & 8,10  \\
UDF12-4071-7347	 & 3:32:40.71	 & -27:47:34.7	 &  $<$ 31.4	 & 29.7 $\pm$ 0.2	 & 30.0 $\pm$ 0.4	 & 29.4 $\pm$ 0.2	 & 30.0 $\pm$ 0.6	 & 7,10  \\
UDF12-4036-8022	 & 3:32:40.36	 & -27:48:02.2	 &  $<$ 31.2	 & 30.1 $\pm$ 0.2	 & 30.0 $\pm$ 0.4	 & 29.9 $\pm$ 0.3	 & 30.7 $\pm$ 1.3	 & 3,7,10  \\
UDF12-3922-6149	 & 3:32:39.22	 & -27:46:14.9	 & 30.8 $\pm$ 1.0	 & 30.0 $\pm$ 0.2	 & 30.2 $\pm$ 0.5	 & 30.6 $\pm$ 0.6	 & 30.1 $\pm$ 0.3	 & 10  \\
UDF12-4245-6534	 & 3:32:42.45	 & -27:46:53.4	 &  $<$ 30.9	 & 30.3 $\pm$ 0.2	 & 30.2 $\pm$ 0.4	 & 29.5 $\pm$ 0.2	 & 29.6 $\pm$ 0.2	 & 11  \\
UDF12-4263-6416	 & 3:32:42.63	 & -27:46:41.6	 & 31.1 $\pm$ 3.2	 & 30.0 $\pm$ 0.2	 & 30.2 $\pm$ 0.3	 & 30.3 $\pm$ 0.5	 & 30.1 $\pm$ 0.4	 & 10  \\
UDF12-4090-6084	 & 3:32:40.90	 & -27:46:08.4	 &  $<$ 31.3	 & 30.2 $\pm$ 0.3	 & 30.5 $\pm$ 0.6	 & 30.3 $\pm$ 0.4	 & 29.8 $\pm$ 0.3	 & 11  \\
UDF12-4019-6190	 & 3:32:40.19	 & -27:46:19.0	 &  $<$ 31.0	 & 29.9 $\pm$ 0.2	 & 30.5 $\pm$ 0.6	 & 29.9 $\pm$ 0.3	 & 29.8 $\pm$ 0.2	 & 10  \\

  \smallskip
\enddata
\tablecomments{z-drop photometry.  All magnitudes listed are measured in the 70$\%$ inclusive aperture
sizes listed in Section 2.4 and not corrected to total here.  We also note
that errors in magnitude space become significantly asymmetric below $\sim10\sigma$.  We have plotted
the larger error wherever appropriate. Upper limits are 1$\sigma$.  
References are (1) \cite{mclure2010a} (2) \cite{oesch2010a}, (3) \cite{finkelstein2010a},
(4) \cite{wilkins2011a}, (5) \cite{yan2010a} (6) \cite{lorenzoni2011a} (7) \cite{bouwens2011a} 
(8) \cite{bouwens2011a} potential (9) \cite{mclure2011a} (10) \cite{mclure2012a} robust
(11) \cite{mclure2012a} potential }
\tablenotetext{a}{Photometric errors $<10\%$.}

\end{deluxetable*}

\begin{deluxetable*}{lccccccc}
\tablecolumns{8}
\tablewidth{0pt}
\tablecaption{\bf $Y$-drops  $7.3<z<8.5$}
\tablehead{\colhead{ID} & \colhead{RA} & \colhead{Dec}
& \colhead{$Y_{105W}$} & \colhead{$J_{125W}$} 
& \colhead{$J_{140W}$} & \colhead{$H_{160W}$} & \colhead{References}}
\medskip
\startdata

UDF12-3880-7072	 & 3:32:38.80	 & -27:47:07.2	 & 28.0 $\pm$ 0.1\tablenotemark{a}	 & 27.3 $\pm$ 0.1\tablenotemark{a}	 & 27.3 $\pm$ 0.1\tablenotemark{a}	 & 27.2 $\pm$ 0.1\tablenotemark{a}	 & 1,2,3,4,5,7,9,10  \\
UDF12-4470-6443	 & 3:32:44.70	 & -27:46:44.3	 & 28.5 $\pm$ 0.1	 & 27.8 $\pm$ 0.1\tablenotemark{a}	 & 27.8 $\pm$ 0.1\tablenotemark{a}	 & 27.8 $\pm$ 0.1\tablenotemark{a}	 & 1,2,3,5,7,9,10  \\
UDF12-3952-7174	 & 3:32:39.52	 & -27:47:17.4	 & 29.3 $\pm$ 0.1	 & 28.3 $\pm$ 0.1	 & 28.2 $\pm$ 0.1	 & 28.0 $\pm$ 0.1	 & 1,2,3,5,7,9,10  \\
UDF12-4314-6285	 & 3:32:43.14	 & -27:46:28.5	 & 29.0 $\pm$ 0.1	 & 28.3 $\pm$ 0.1	 & 28.2 $\pm$ 0.1	 & 28.1 $\pm$ 0.1\tablenotemark{a}	 & 1,2,3,4,5,7,9  \\
UDF12-3722-8061	 & 3:32:37.22	 & -27:48:06.1	 & 29.0 $\pm$ 0.1	 & 28.3 $\pm$ 0.1	 & 28.3 $\pm$ 0.1	 & 28.3 $\pm$ 0.1	 & 1,2,3,5,7,9,10  \\
UDF12-3813-5540	 & 3:32:38.13	 & -27:45:54.0	 & 30.1 $\pm$ 0.2	 & 28.6 $\pm$ 0.1	 & 28.5 $\pm$ 0.1	 & 28.4 $\pm$ 0.1	 & 1,3,5,6,7,9,10  \\
UDF12-3780-6001	 & 3:32:37.80	 & -27:46:00.1	 & 29.8 $\pm$ 0.2	 & 28.6 $\pm$ 0.1	 & 28.6 $\pm$ 0.1	 & 28.7 $\pm$ 0.1	 & 1,3,5,6,7,10  \\
UDF12-3764-6015	 & 3:32:37.64	 & -27:46:01.5	 & 30.5 $\pm$ 0.3	 & 28.9 $\pm$ 0.1	 & 28.9 $\pm$ 0.1	 & 28.8 $\pm$ 0.1	 & 1,3,5,7,10  \\
UDF12-3939-7040	 & 3:32:39.39	 & -27:47:04.0	 & 30.0 $\pm$ 0.2	 & 29.1 $\pm$ 0.1	 & 29.1 $\pm$ 0.1	 & 28.9 $\pm$ 0.1	 & 10  \\
UDF12-4474-6450	 & 3:32:44.74	 & -27:46:45.0	 & 29.7 $\pm$ 0.2	 & 29.0 $\pm$ 0.1	 & 29.1 $\pm$ 0.2	 & 29.1 $\pm$ 0.1	 & 1,2,3,5,7,10  \\
UDF12-4309-6277	 & 3:32:43.09	 & -27:46:27.7	 & 30.3 $\pm$ 0.3	 & 28.9 $\pm$ 0.1	 & 29.4 $\pm$ 0.2	 & 29.3 $\pm$ 0.1	 & 1,3,5,7,11  \\
UDF12-4309-6260	 & 3:32:43.09	 & -27:46:26.0	 & 30.1 $\pm$ 0.2	 & 29.3 $\pm$ 0.2	 & 28.9 $\pm$ 0.1	 & 29.3 $\pm$ 0.2	 & 7,10  \\
UDF12-3463-6472	 & 3:32:34.63	 & -27:46:47.2	 & 30.3 $\pm$ 0.2	 & 29.6 $\pm$ 0.2	 & 29.3 $\pm$ 0.2	 & 29.4 $\pm$ 0.2	 & 7  \\
UDF12-3551-7443	 & 3:32:35.51	 & -27:47:44.3	 & 30.7 $\pm$ 0.4	 & 29.5 $\pm$ 0.2	 & 29.9 $\pm$ 0.4	 & 29.4 $\pm$ 0.3	 & 7,11  \\
UDF12-4336-6203	 & 3:32:43.36	 & -27:46:20.3	 &  $<$ 31.8	 & 29.5 $\pm$ 0.2	 & 29.6 $\pm$ 0.2	 & 29.4 $\pm$ 0.2	 & -  \\
UDF12-4240-6550	 & 3:32:42.40	 & -27:46:55.0	 & 30.4 $\pm$ 0.3	 & 29.4 $\pm$ 0.1	 & 29.7 $\pm$ 0.2	 & 29.5 $\pm$ 0.2	 & 7,10  \\
UDF12-4033-8026	 & 3:32:40.33	 & -27:48:02.6	 & 30.1 $\pm$ 0.3	 & 29.3 $\pm$ 0.2	 & 29.6 $\pm$ 0.3	 & 29.5 $\pm$ 0.3	 & 3,7,10  \\
UDF12-4308-6242	 & 3:32:43.08	 & -27:46:24.2	 & 30.4 $\pm$ 0.3	 & 29.7 $\pm$ 0.3	 & 30.0 $\pm$ 0.4	 & 29.6 $\pm$ 0.2	 & 7,10  \\
UDF12-3931-6180	 & 3:32:39.31	 & -27:46:18.0	 & 29.8 $\pm$ 0.2	 & 29.3 $\pm$ 0.2	 & 29.3 $\pm$ 0.1	 & 29.6 $\pm$ 0.2	 & 10  \\
UDF12-3934-7256	 & 3:32:39.34	 & -27:47:25.6	 & 31.2 $\pm$ 0.7	 & 29.8 $\pm$ 0.3	 & 30.0 $\pm$ 0.3	 & 29.7 $\pm$ 0.2	 & 7,11  \\
UDF12-3881-6343	 & 3:32:38.81	 & -27:46:34.3	 & 30.6 $\pm$ 0.4	 & 29.9 $\pm$ 0.4	 & 30.0 $\pm$ 0.4	 & 29.8 $\pm$ 0.3	 & 11  \\
UDF12-3681-6421	 & 3:32:36.81	 & -27:46:42.1	 &  $<$ 31.4	 & 30.0 $\pm$ 0.4	 & 30.0 $\pm$ 0.4	 & 29.8 $\pm$ 0.3	 & 7,11  \\
UDF12-4294-6560	 & 3:32:42.94	 & -27:46:56.0	 & 30.8 $\pm$ 0.6	 & 29.9 $\pm$ 0.3	 & 30.5 $\pm$ 0.7	 & 29.9 $\pm$ 0.3	 & 11  \\
UDF12-3920-6322	 & 3:32:39.20	 & -27:46:32.2	 &  $<$ 31.9	 & 29.9 $\pm$ 0.3	 & 29.7 $\pm$ 0.3	 & 29.9 $\pm$ 0.3	 & 10  \\
UDF12-3858-6150	 & 3:32:38.58	 & -27:46:15.0	 & 30.3 $\pm$ 0.3	 & 29.8 $\pm$ 0.3	 & 30.3 $\pm$ 0.4	 & 29.9 $\pm$ 0.3	 & 11  \\
UDF12-4344-6547	 & 3:32:43.44	 & -27:46:54.7	 &  $<$ 31.9	 & 30.0 $\pm$ 0.3	 & 30.1 $\pm$ 0.4	 & 30.0 $\pm$ 0.3	 & 10  \\
UDF12-4101-7216	 & 3:32:41.01	 & -27:47:21.6	 & 30.4 $\pm$ 0.2	 & 29.9 $\pm$ 0.3	 & 30.0 $\pm$ 0.3	 & 30.9 $\pm$ 0.8	 & 10  \\

\smallskip
\enddata
\tablecomments{Y-drop photometry.  All magnitudes listed are measured in the 70$\%$ inclusive aperture
sizes listed in Section 2.4 and not corrected to total here.  We also note
that errors in magnitude space become significantly asymmetric below $\sim10\sigma$.  We have plotted
the larger error wherever appropriate.  Upper limits are 1$\sigma$.  
References are (1) \cite{mclure2010a} (2) \cite{oesch2010a}, (3) \cite{finkelstein2010a},
(4) \cite{wilkins2011a}, (5) \cite{yan2010a} (6) \cite{lorenzoni2011a} (7) \cite{bouwens2011a} 
(8) \cite{bouwens2011a} potential (9) \cite{mclure2011a} (10) \cite{mclure2012a} robust
(11) \cite{mclure2012a} potential }
\tablenotetext{a}{Photometric errors $<10\%$.}

\end{deluxetable*}

\begin{deluxetable*}{lc}
\tablecolumns{2}
\tablewidth{0pt}
\tablecaption{\bf SWML determination of the $z\sim7$ LF}
\tablehead{\colhead{$M_{\mathrm{UV}}$} & \colhead{log $\phi_k$ [Mpc$^{-3}$ mag$^{-1}$]}}
\medskip
\startdata
-20.65  &  -4.29$^{+0.29}_{-0.28}$ \\
-20.15  &  -3.71$^{+0.14}_{-0.10}$ \\
-19.65  &  -3.31$^{+0.08}_{-0.10}$ \\
-19.15  &  -3.02$^{+0.13}_{-0.06}$ \\
-18.65  &  -2.98$^{+0.17}_{-0.23}$ \\
-18.15  &  -2.56$^{+0.19}_{-0.06}$ \\
-17.65  &  -2.23$^{+0.12}_{-0.09}$ \\
-17.15  &  -3.03$^{+0.54}_{-2.34}$ \\
 \smallskip
\enddata
\end{deluxetable*}

\begin{deluxetable*}{lc}
\tablecolumns{2}
\tablewidth{0pt}
\tablecaption{\bf SWML determination of the $z\sim8$ LF}
\tablehead{\colhead{$M_{\mathrm{UV}}$} & \colhead{log $\phi_k$ [Mpc$^{-3}$ mag$^{-1}$]}}
\medskip
\startdata
-22.00  &  $<5.01$ \\
-21.50  &  -5.02$^{+0.44}_{-0.47}$ \\
-21.00  &  -4.28$^{+0.16}_{-0.24}$ \\
-20.50  &  -4.15$^{+0.12}_{-0.43}$ \\
-20.00  &  -3.54$^{+0.17}_{-0.06}$ \\
-19.50  &  -3.34$^{+0.15}_{-0.17}$ \\
-19.00  &  -2.97$^{+0.09}_{-0.20}$ \\
-18.50  &  -2.91$^{+0.14}_{-0.24}$ \\
-18.00  &  -2.61$^{+0.18}_{-0.20}$ \\
-17.50  &  -2.57$^{+0.25}_{-0.74}$ \\
 \smallskip
\enddata
\end{deluxetable*}

\begin{figure*}
\begin{center}
\includegraphics[width=0.4\textwidth]{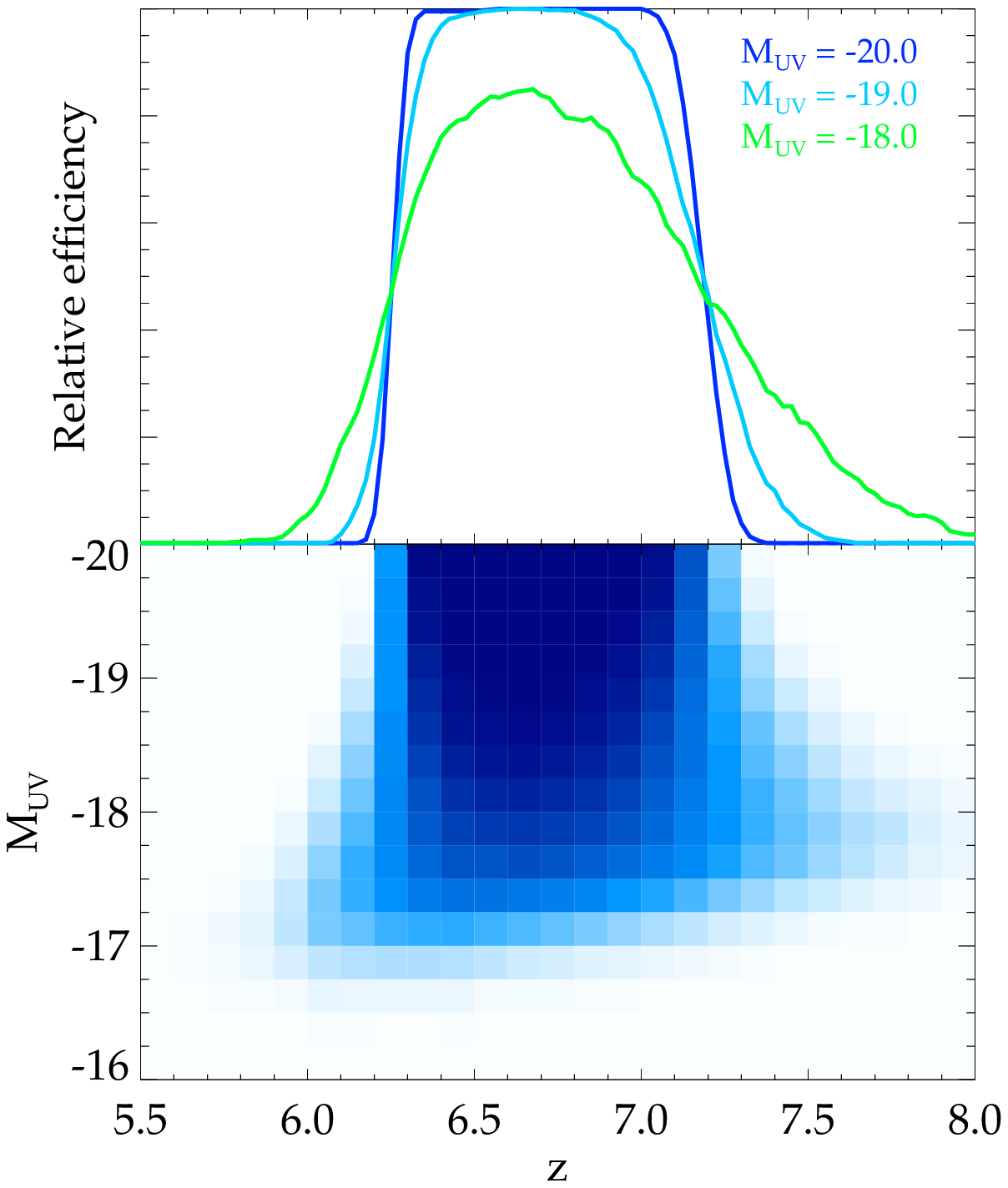}
\includegraphics[width=0.4\textwidth]{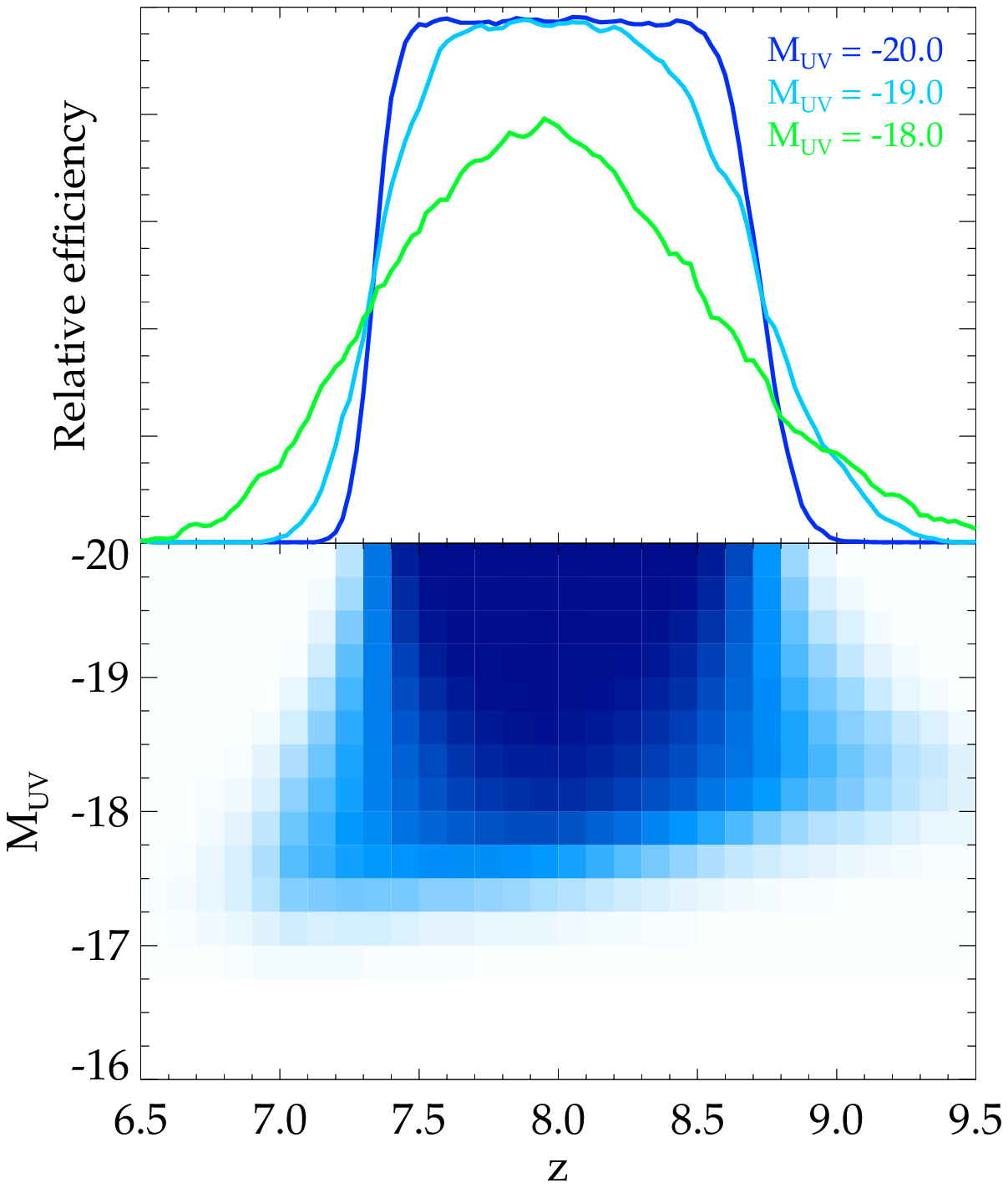}
\caption{Left: Selection function for $z_{850}$-drop galaxies in the UDF, as a function of $M_{\mathrm{UV}}$ and 
redshift, constructed using the simulations discussed in section 4.1.  Even at bright magnitudes, 
maximum efficiency is only $\sim$0.65 due to the area subtended by other objects and our strict optical non-detection
criteria, which result in a small fraction of true high-redshift sources being excluded.  Right: Equivalent selection function for
$Y_{105}$-drops.}
\end{center}
\end{figure*}

\begin{figure*}
\begin{center}
\includegraphics[width=0.45\textwidth]{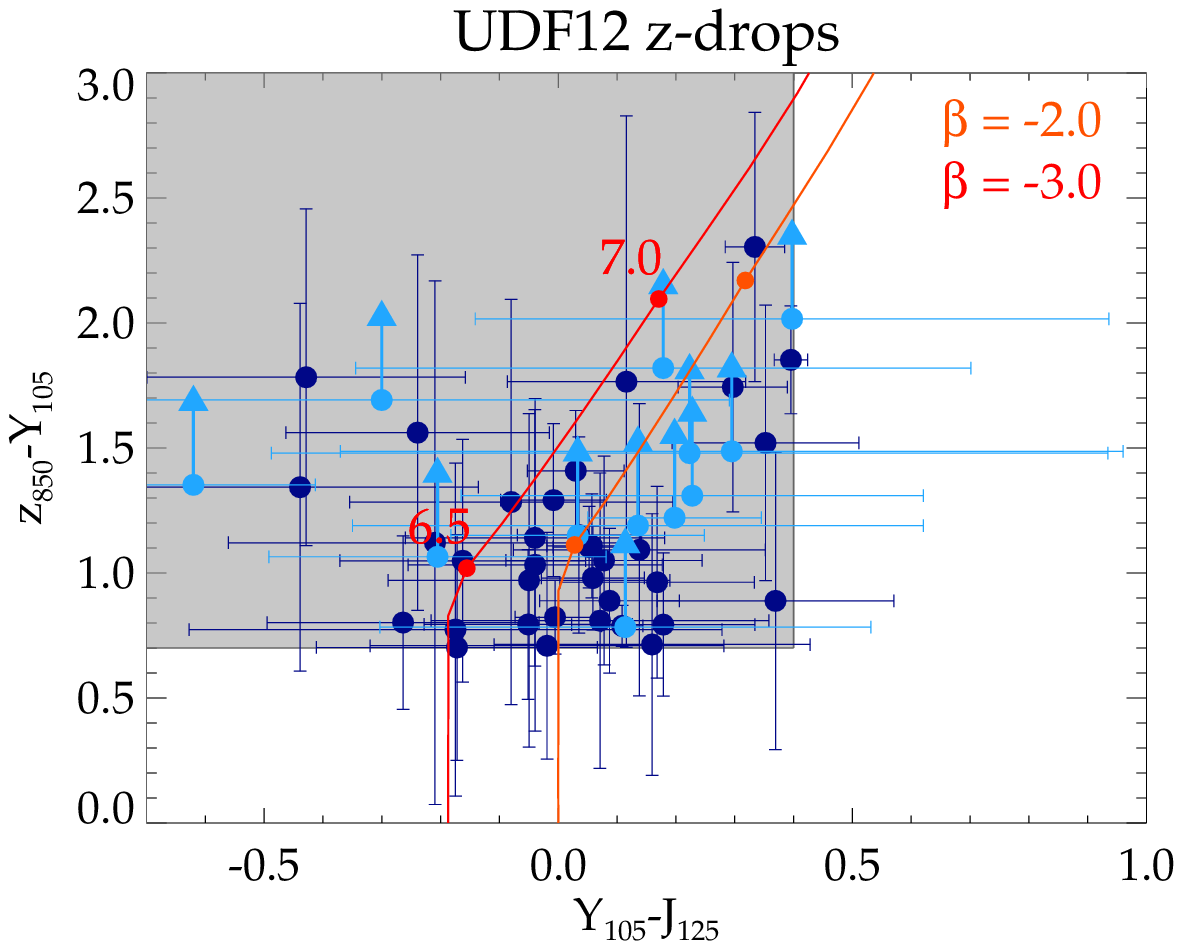}
\includegraphics[width=0.45\textwidth]{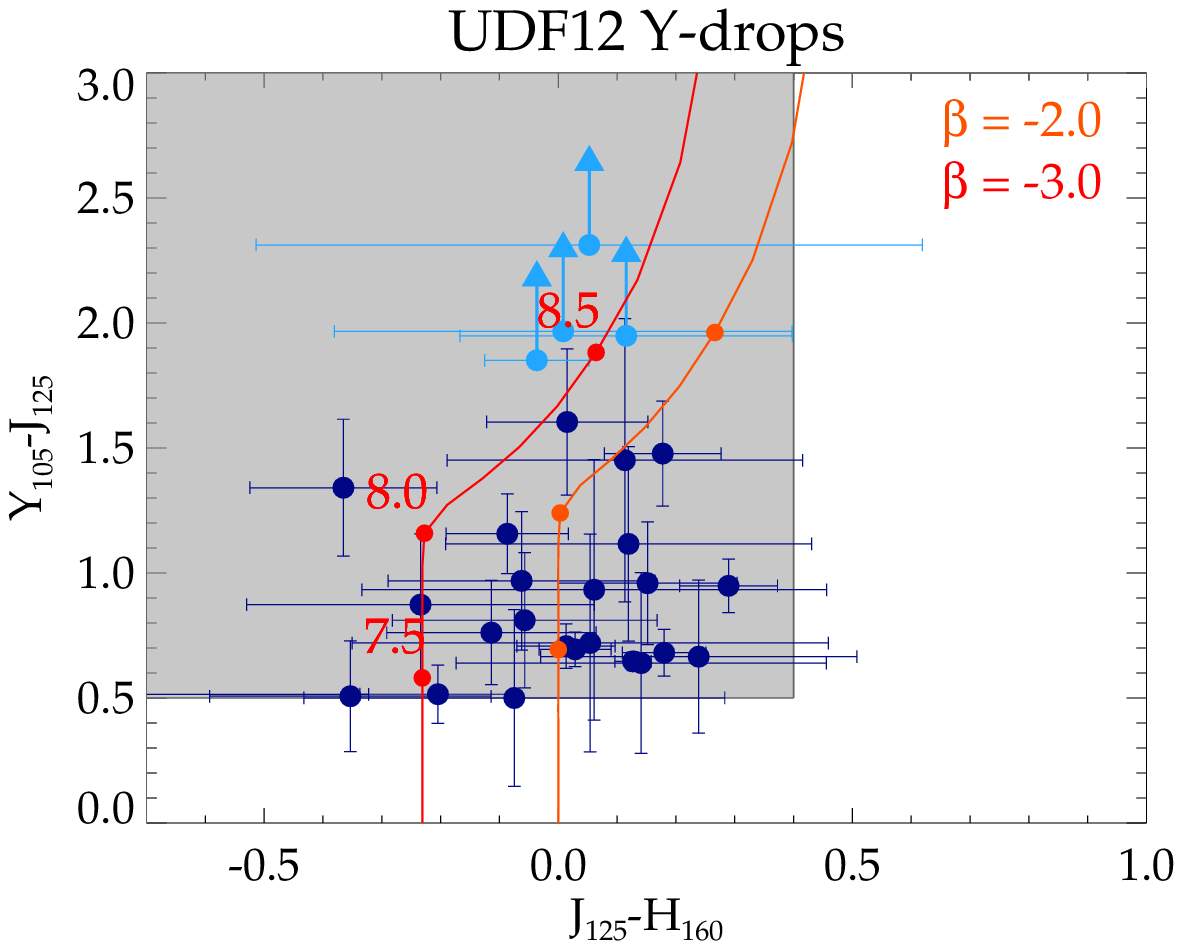}
\caption{\label{fig:cand_plot}Color-color diagram of galaxies selected as $z_{850}$-drops (left) and 
$Y_{105}$-drops (right).  Red and 
orange curves show the tracks of synthetic high-redshift galaxies for various UV continuum slopes $\beta$.  
Light blue points denote 1-$\sigma$ upper limits.}
\end{center}
\end{figure*} 

\begin{figure*}
\begin{center}
\includegraphics[width=0.6\textwidth]{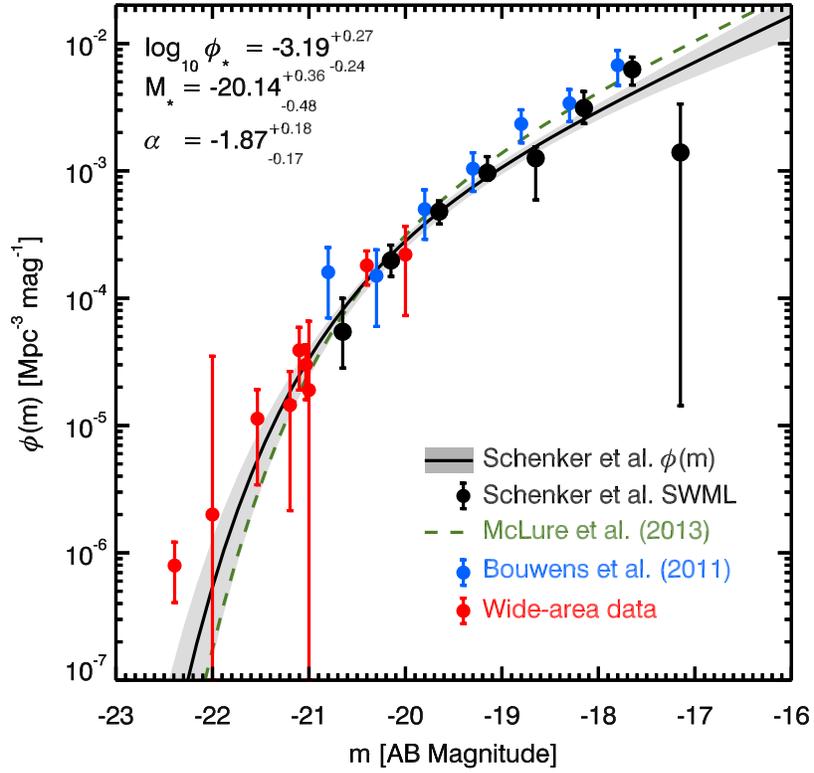}
\caption{\label{fig:z7_lf}The luminosity function of star-forming galaxies at $z \sim 7$ from the $z_{850}$-drop sample.  Black points 
were determined using the UDF12 data set and other HST data mentioned in this work.  Red points denote wide area ground based data
increase the range in luminosity. The black line defines the maximum likelihood Schechter luminosity function and the shaded grey 
region denotes the 68$\%$ confidence interval. The green dashed line denotes the fit of \cite{mclure2012a}.}
\end{center}
\end{figure*} 

\begin{figure*}
\begin{center}
\includegraphics[width=0.6\textwidth]{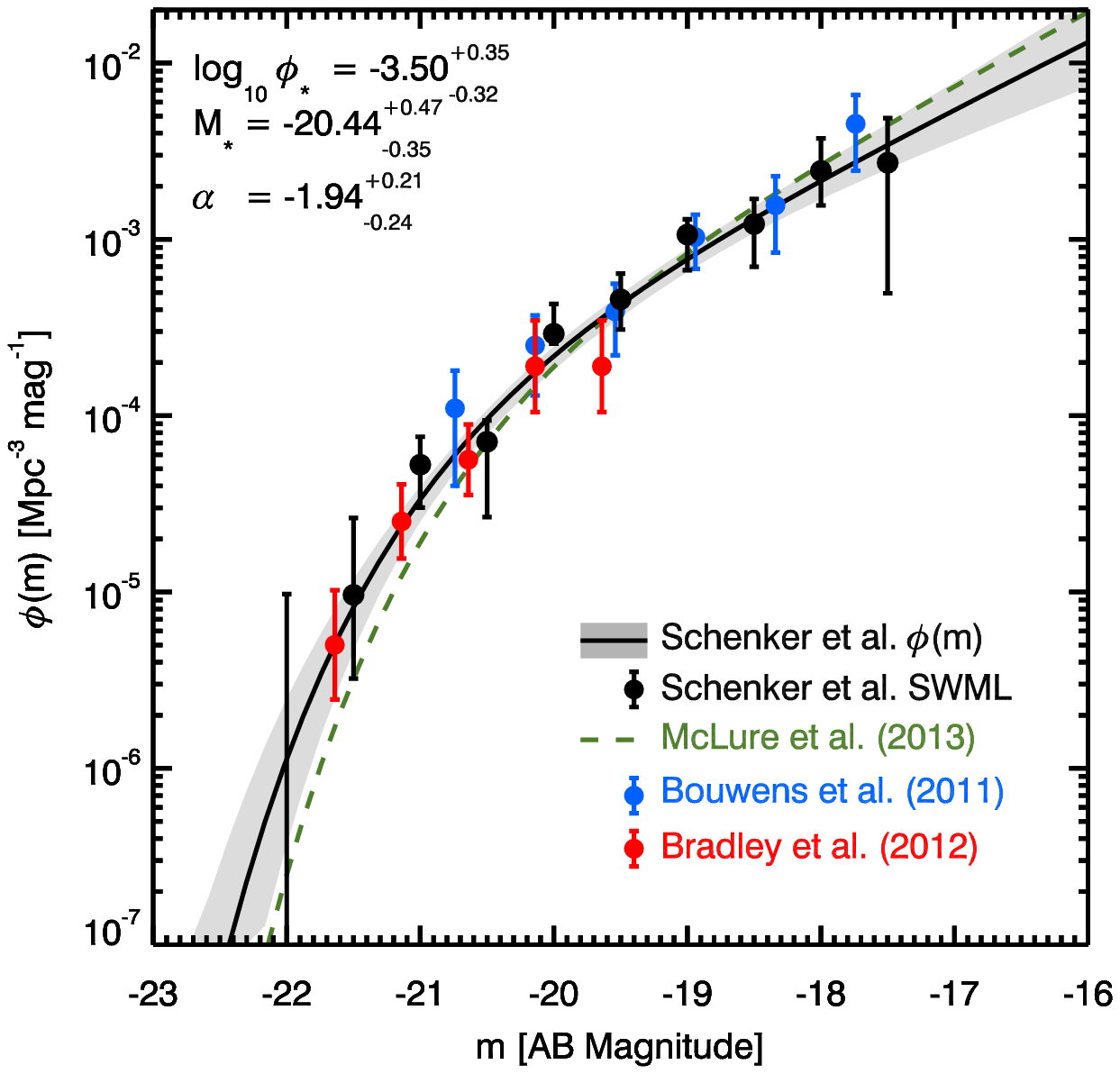}
\caption{\label{fig:z8_lf}The luminosity function of star-forming galaxies at $z \sim 8$ from the $Y_{105}$-drop sample.  Black points 
were determined using the  UDF12 data set and other HST data mentioned in this work.  Red points denote data from the 
\cite{bradley2012a} analysis of the BoRG fields that increase the range in luminosity.  The black line
defines the maximum likelihood Schechter luminosity function and the shaded grey region denotes the 68$\%$ confidence
interval.  The green line denotes our fit when removing the BoRG dataponts; we note that our fit to the faint
end slope is remarkably insensitive to their inclusion/exclusion.  The green dashed line denotes the fit of \cite{mclure2012a}. }
\end{center}
\end{figure*}

\end{document}